\documentclass[a4paper]{iopart}
\pdfoutput=1
\usepackage{textcomp}
\usepackage{amsmath}
\usepackage{amssymb}
\usepackage{graphicx} 
\usepackage{pgf}
\usepackage[showframe=false]{geometry}
\usepackage{mathdots}
\usepackage[pdftex,bookmarks,colorlinks]{hyperref}
\usepackage{bbold}

\newcommand{\Trace}{{\mathrm{Tr}}}
\def\bra#1{\langle#1 |}
\def\ket#1{| #1\rangle}

\def\ud{\mathrm{d}}

\newcommand{\nep}{\textrm{e}}
\def\op#1{\hat{#1}}

\newcommand{\bsigma}{\boldsymbol{\sigma}}
\newcommand{\bpauli}{\hat{\boldsymbol{\sigma}}}
\newcommand{\bfb}{{\bf b}}

\newcommand{\bfr}{{\bf r}}

\newcommand{\n}{{\bf n}}

\newcommand{\R}{{\bf R}}

\newcommand{\mathA}{{\mathbb{A}}}

\newcommand{\opadag}[1]{{\hat{a}^{\dagger}}_{#1}}
\newcommand{\opa}[1]{{\hat{a}^{\phantom \dagger}}_{#1}}
\newcommand{\opbdag}[1]{{\hat{b}^{\dagger}}_{#1}}
\newcommand{\opb}[1]{{\hat{b}^{\phantom \dagger}}_{#1}}
\newcommand{\opcdag}[1]{{\hat{c}^{\dagger}}_{#1}}
\newcommand{\opc}[1]{{\hat{c}^{\phantom \dagger}}_{#1}}

\newcommand{\pauli}{\op{\sigma}}

\newcommand{\hc}{\mathrm{H.c.}}

\newcommand{\id}{\mathbb{1}}
\newcommand{\ie}{\textit{i.e.} }
\newcommand{\vs}{\textit{vs} }

\newcommand{\coh}{\mathrm{coh}}

\newcommand{\Ham}{\widehat{H}}
\newcommand{\BZ}{\mathrm{BZ}}
\newcommand{\Hsys}{\Ham_S}

\newcommand{\Hcalsysk}{\hat{\mathcal{H}}_S^k}
\newcommand{\HSB}{\Ham_{SB}}

\newcommand{\Hbath}{\Ham_B}

\newcommand{\Htot}{\Ham_{\mathrm{tot}}}

\newcommand{\rhosys}{\hat{\rho}_{\scriptscriptstyle \mathrm{S}}}
\newcommand{\rhosysk}{\hat{\rho}_{\scriptscriptstyle \mathrm{S}}^{k}}
\newcommand{\rhoFlqt}[1]{\rho_{#1}^k}
\newcommand{\rhoFlqtcoh}[1]{\rho_{#1}^{k,\coh}}

\newcommand{\period}{\tau}
\newcommand{\tmpr}{T}
\newcommand{\Jop}{\hat{\mathrm J}}
\newcommand{\Jkt}{\hat{\mathcal{J}}}
\newcommand{\Qinfty}{\overline{Q}}
\newcommand{\Qdiag}{Q_{\rm d}^{\coh}}
\newcommand{\Qdiagdiss}{Q_{\rm d}^{\rm diss}}
\newcommand{\Fstate}{\psi}
\newcommand{\Fmode}{u}
\newcommand{\Fqe}{\epsilon}

\newcommand{\Gpuredeph}{\gamma_\varphi}
\newcommand{\Grelax}{\gamma_{\scriptscriptstyle \mathrm{R}}}
\newcommand{\Asmall}{\mathrm{\scriptscriptstyle A}}
\newcommand{\Bsmall}{\mathrm{\scriptscriptstyle B}}

\newcommand{\GdephRWA}[1]{\gamma_{{\scriptscriptstyle \mathrm{D}},#1}}
\newcommand{\gammaRWA}{\gamma}
\newcommand{\GdephNORWA}{\tilde{\gamma}_{\scriptscriptstyle \mathrm{D}}}
\newcommand{\gammaNORWA}{\tilde{\gamma}}
\newcommand{\transpose}{\scriptscriptstyle\mathrm{T}}

\begin{document}

\title{Dissipation assisted Thouless pumping in the Rice-Mele model}

\author{Luca Arceci$^{1}$, Lucas Kohn$^{1}$, Angelo Russomanno$^{2,3}$, Giuseppe E. Santoro$^{1,2,4}$}

\address{$^1$ SISSA, Via Bonomea 265, I-34136 Trieste, Italy}
\address{$^2$ International Centre for Theoretical Physics (ICTP), P.O. Box 586, I-34014 Trieste, Italy}
\address{$^3$ Max-Planck-Institut f\"ur Physik Komplexer Systeme, N\"othnitzer Strasse 38, D-01187, Dresden, Germany}
\address{$^4$ IOM-CNR Democritos c/o SISSA, Via Bonomea 265, I-34136 Trieste, Italy}
%\address{$^2$ ICTP, Strada Costiera 11, 34151 Trieste, Italy}

\eads{\mailto{larceci@sissa.it}, \mailto{arussoma@ictp.it}, \mailto{santoro@sissa.it}}
%
%\pacs{75.10.Pq, 05.30.Rt, 03.65.-w}

\begin{abstract}
We investigate the effect of dissipation from a thermal environment on topological pumping in the periodically-driven Rice-Mele model. 
We report that dissipation can improve the robustness of pumping quantisation in a regime of finite driving frequencies. 
Specifically, in this regime, a low-temperature dissipative dynamics can lead to a pumped charge that is much closer to the Thouless quantised value, 
compared to a coherent evolution. 
We understand this effect in the Floquet framework: dissipation increases the population of a Floquet band which shows a topological winding, where pumping is essentially quantised. 
This finding is a step towards understanding a potentially very useful resource to exploit in experiments, where dissipation effects are unavoidable. 
We consider small couplings with the environment and we use a Bloch-Redfield quantum master equation approach for our numerics: Comparing these results with an exact MPS numerical treatment we find that the quantum master equation works very well also at low temperature, a quite remarkable fact.
\end{abstract}

\maketitle

%----------------------------------------------
\section{Introduction}		\label{sec:intro}
%----------------------------------------------
%
Quantised adiabatic transport in insulators, discovered by Thouless \cite{Thouless_PRB83} in 1983, is a topic of current
interest in the field of topological insulators, recently fostered by the experimental realization of a Thouless pump with 
ultra-cold atoms \cite{Nakajima_NatPhys16,Bloch_NatPhys16}.
%
%in topological insulators is an interesting topic in current research in condensed matter physics. 
%Firstly discovered by Thouless \cite{Thouless_PRB83}, the field has been rapidly growing in the last years, 
%recently fostered by the first experimental realization of a Thouless pump with ultra-cold atoms \cite{Bloch_NatPhys16}. 
%This last achievement raises the hope for new quantum technologies in the near future.
%

The strict quantisation of the pumped charge requires the quantum dynamics to be, in principle, adiabatic and unitary. 
However, in concrete experimental realizations these two requirements cannot be perfectly fulfilled. 
The study of non-adiabatic effects on a specific example of Thouless pump, the so-called Rice-Mele model \cite{RiceMele_PRL82}, 
showed the emergence of quadratic corrections in the driving frequency whenever the system is initially prepared in the initial Hamiltonian 
ground state \cite{Privitera_PRL18}. 
Similar non-adiabatic effects have been discussed in the context of the quantisation of the Hall conductivity in the Harper-Hofstadter model \cite{Wauters_PRB18}. 
The effect of disorder on the topological pumping of the Rice-Mele model has been recently discussed in Ref.~\cite{Wauters_PRL20}. 
An important point is also the effect of interactions, which make the system non integrable and lead the system to heat up $\tmpr = \infty$ in the long-time limit. 
In this case, adiabatic pumping is asymptotically washed out and can occur only as a transient condition \cite{Lindner_PRX17}. 

Dissipative effects may also disrupt pumping quantisation. 
However, the impact and role of dissipation in the performance of a Thouless pump remains an open question. 
Some studies have taken into account thermal effects by using a thermal initial state, instead of the Hamiltonian ground state 
\cite{Wang_PRL13, Privitera_PRL18}, followed by a unitary dynamics. 
Within such a framework, it was found~\cite{Wang_PRL13} that charge quantisation is robust against non-zero temperatures in the initial state when a single pumping 
cycle is considered. Moreover, in the limit of an infinite number of pumping cycles \cite{Privitera_PRL18}, 
thermal corrections were found to be exponentially small for low enough temperatures but increasingly relevant when the temperature approached the insulating gap. 
Concerning papers where a genuine dissipative dynamics is considered, we mention a study \cite{Zhou_PRB15} of the Qi-Wu-Zhang model through a 
Lindblad Markovian quantum master equation, where it is found that the pumped charge, starting from the quantisation value, decreases monotonically to zero with
increasing noise. 
 
%With increasing noise, it is found that the pumped charge starts from a value equal to 1 and decreases monotonically to zero. 
%\BLUE{In the simpler setting of non-topological pumping in a three-sites fermionic chain, dissipation from a thermal reservoir has also been studied \cite{Pellegrini_PRL11}}.

In this work, we aim at understanding how topological quantum pumping is affected by the interaction with a bosonic thermal bath. 
To do so, we study the time evolution of the Rice-Mele Hamiltonian \cite{RiceMele_PRL82} and analyse how charge pumping is affected 
both by dissipation and by non-adiabatic effects. 
The remarkable finding of our study is that --- in appropriate conditions, \ie if the temperature of the bath is low enough, and in the limit of infinite cycles ---
a dissipative dynamics may be {\em beneficial} to a Thouless pump, fighting against non-adiabatic effects, and leading to a pumped charge which is closer to
the quantised value. 
%
%\st{That is, at fixed driving period, the charge pumped in the dissipative time evolution is closer to the quantised value, 
%as compared to the one coming from purely coherent dynamics. }

We rationalise our findings by analysing the dissipative results in terms of the Floquet states of the unitary dynamics. 
Interestingly, we find that the charge pumped at stationarity is still expressed, exactly as in the unitary case, only in terms of the populations of 
the Floquet bands, with no role for the quantum coherences. 
Quantised pumping is essentially related to the perfect occupation of what one might call the ``lowest-energy Floquet band'', \ie the Floquet band 
constructed by choosing, for each momentum, the Floquet mode with (period-averaged) lowest-energy expectation: 
this is the quasi-energy band ``closest'' to the instantaneous ground state, that shows the correct {\em winding} \cite{kitagawa2010topological} 
across the Brillouin zone (BZ) and leads to a quantised charge transport, 
up to non-adiabatic corrections which are exponentially small in the driving period, see Refs.~\cite{Shih_PRB94,Privitera_PRL18}.
The dissipation-induced improved pumping is found to be strictly related to the fact that dissipation brings to a higher population for such a lowest-energy Floquet band. 
We might regard it as a dissipative preparation of a topological Thouless pump, somewhat similar in spirit to the results of Ref.~\cite{Budich_PRA15},
which however deals with a two-dimensional non-driven system with a specifically engineered dissipation.
It is also worth to mention Ref.~\cite{Pekola_PRL10}, dealing with non-topological pumping in superconducting nanocircuits. The authors show that a zero-temperature bath can make the pumped charge at finite frequency closer to its coherent adiabatic counterpart. This phenomenon occurs also in our case. 

The paper is organised as follows: 
in Sec.~\ref{sec:model} we introduce our dissipative version of the Rice-Mele model. 
In Sec.~\ref{sec:charge} we define the pumped charge, \ie the observable displaying quantisation in the non-dissipative adiabatic limit, 
and we describe it within the Floquet theory. 
%We furthermore introduce  and establish a connection between the pumped charge at stationarity and the Floquet diagonal ensemble. 
In Sec.~\ref{sec:methods} we illustrate how we compute the dissipative time evolution of the system. 
In particular, we benchmark the reliability of the Bloch-Redfield Quantum Master Equation against a non-perturbative approach, which exploits a unitary chain-mapping transformation in order to carry out time evolution with Matrix Product States. We find a very good agreement, especially at T=0 where the quantum master equation was not expected to work~\cite{Weiss:book}.
%two approaches for computing the non-unitary time evolution of the system: 
%the Bloch-Redfield Quantum Master Equation (QME), 
%very fast and versatile but subject to approximations, 
%which is used to obtain all the results in this work
%The second is an approach based on the so-called chain mapping and time evolution through Matrix Product States (MPS) and the Time Dependent Variational Principle (TDVP). This method is computationally heavier but is not subject to all the approximations of the QME. We therefore use it to benchmark the QME results. 
%\newline
%(PUT CITATIONS)
%\newline

In Sec. \ref{sec:results} we present our numerical results for the pumped charge and discuss the pumping enhancement due to a low-temperature bath. 
We deepen the understanding of this effect in the Floquet framework, which provides insights in terms of populations of the Floquet bands. 
Finally, in Sec. \ref{sec:conclusions} we summarize the main results and draw our conclusions. 

%------------------------------------------
\section{The model}		\label{sec:model}
%------------------------------------------
%
We study here a dissipative version of the Rice-Mele model, where the non-unitary dynamics comes from coupling the system to a bosonic bath at thermal equilibrium. 
The Hamiltonian is
\begin{equation}	\label{H_tot}
	\Htot(t) = \Hsys(t) + \HSB + \Hbath
\end{equation}
where the three terms on the r.h.s. are, respectively, the system, system-bath interaction and bath Hamiltonians.

The system Hamiltonian consists of a bipartite lattice on which spinless fermions hop on nearest-neighbor sites, according to the Rice-Mele Hamiltonian \cite{RiceMele_PRL82}:
\begin{equation}	\label{H_RM}
	\Hsys(t) = - \sum_{j=1}^N \Big( J_1(t) \opcdag{j,\Bsmall} \opc{j,\Asmall} + J_2(t) \opcdag{j+1,\Asmall} \opc{j,\Bsmall} + \hc \Big) 
	+ \Delta(t) \, \sum_{j=1}^N \Big( \opcdag{j,\Asmall}\opc{j,\Asmall} - \opcdag{j,\Bsmall}\opc{j,\Bsmall} \Big)
\end{equation}
where $N$ is the number of diatomic cells,
$\opcdag{j,\Asmall(\Bsmall)}$ creates a fermion on site $A(B)$ of the $j^{th}$ cell, 
$J_1(t)$ and $J_2(t)$ are respectively intra-cell and inter-cell hopping terms and $\Delta(t)$ modulates the on-site energies. 
We assume periodic boundary conditions (PBC), 
so that translational invariance allows us to Fourier transform the fermionic operators for the $A$ and $B$ sites separately, 
$\opc{j,\Asmall(\Bsmall)} = \sum_{k} e^{ikaj} \opc{k,\Asmall(\Bsmall)} / \sqrt{N}$, 
where the sum over the discrete wave-vectors $k = 2\pi n / (Na)$, with $n = 0, 1, \dots, N-1$ and $a$ the cell length, 
runs inside the first BZ.
%$BZ = \{ k_n = 2\pi n / N, \; n = 0, 1, \dots, N-1 \}$.
By applying this transformation to Eq.~\eqref{H_RM}, we block-diagonalise the Hamiltonian in sectors of different $k$: 
%$\Hsys(t) = \sum_{k \in BZ} \Hsysk(t)$, where:
\begin{equation}	\label{H_RMdiag}
	\Hsys(t) = \sum_{k}^{\BZ}
	\begin{bmatrix}
		\opcdag{k,\Asmall} & \opcdag{k,\Bsmall}
	\end{bmatrix}
	\Hcalsysk(t)
	\begin{bmatrix}
		\opc{k,\Asmall} \\ \opc{k,\Bsmall}
	\end{bmatrix}	
	= \sum_{k}^{\BZ}
	\begin{bmatrix}
		\opcdag{k,\Asmall} & \opcdag{k,\Bsmall}
	\end{bmatrix}
	 \R(k,t) \cdot \bpauli 
	\begin{bmatrix}
		\opc{k,\Asmall} \\ \opc{k,\Bsmall}
	\end{bmatrix}
\end{equation}
where $\Hcalsysk(t) = \R(k,t) \cdot \bpauli$ is a 2-dimensional operator, conveniently parameterised, using the Bloch-vector notation, 
with the vector of Pauli matrices $\bpauli$ and an effective magnetic field $\R(k,t)$
\begin{equation}
\R(k,t) = (-J_1(t) -J_2(t) \cos (ka), -J_2(t) \sin (ka), \Delta(t))^{\transpose} \;.
\end{equation}
% = (-J_1(t) -J_2(t) \cos (ka), -J_2(t) \sin (ka), \Delta(t))$.
%\begin{subequations}	\label{RM_Rcomponents}
%\begin{align}
%	R_x(k,t) &= -J_1(t) -J_2(t) \cos k \\
%	R_y(k,t) &= -J_2(t) \sin k \\
%	R_z(k,t) &= \Delta(t)
%\end{align}
%\end{subequations}
The energy bands $E_\pm(k,t) = \pm |\R(k,t)|$ of $\Hcalsysk(t)$ never ``touch'' except at a single 
``metallic'' point with $\Delta = 0$ and $J_1 = J_2$, where $\R=0$ (for $k = \pi/a$). 
We completely fill the lowest band with $N$ particles --- the half-filling condition --- and realise a quantised adiabatic pumping 
by driving the band insulator around the metallic point with a schedule 
$J_1(t) = J_0 + \delta_0 \cos \omega t$, $J_2(t) = J_0 - \delta_0 \cos \omega t$ and $\Delta(t) = \Delta_0 \sin \omega t$.
%in the parameter space ($J_1(k,t) - J_2(k,t)$, $\Delta(t)$) around the metallic point ($0, 0$).
%Therefore, we parametrize our external fields as follows:
%\begin{subequations}
%\begin{align}
%	J_1(t) &= J_0 + \delta_0 \cos \omega t \\
%	J_2(t) &= J_0 - \delta_0 \cos \omega t \\
%	\Delta(t) &= \Delta_0 \sin \omega t 
%\end{align}
%\end{subequations}
Here $\omega = 2\pi / \period$ is the driving frequency, associated to a period $\period$, so that $\Hsys(t + \period) = \Hsys(t)$. 

%%%%%%%%%%%%%%%%%%%%%%%%%%%%%%%%%%%%
%				BATH
%%%%%%%%%%%%%%%%%%%%%%%%%%%%%%%%%%%%
To account for dissipation in the simplest and most practical way, we choose to include identical but independent harmonic baths
for each $k$-subsector, coupled in the usual Caldeira-Leggett spin-boson~\cite{Leggett_RMP87} fashion
\begin{eqnarray} \label{Htot_k}
\Htot(t)  &= & \sum_{k}^{\BZ} 
\begin{bmatrix}
		\opcdag{k,\Asmall} & \opcdag{k,\Bsmall}
	\end{bmatrix}
	\hat{\mathcal{H}}_k(t)
	\begin{bmatrix}
		\opc{k,\Asmall} \\ \opc{k,\Bsmall}
	\end{bmatrix}	\\
\hat{\mathcal{H}}_k(t) &=&  \R(k,t) \cdot \bpauli + (\n \cdot \bpauli) \otimes \hat{X}_k + \sum_l \hbar\omega_l \opbdag{k,l} \opb{k,l} 
\nonumber
\end{eqnarray}
%
%\begin{align} \label{Htot_k}
%\begin{split}
%\Htot(t)  &=  \sum_{k}^{\BZ} 
%\begin{bmatrix}
%		\opcdag{k,\Asmall} & \opcdag{k,\Bsmall}
%	\end{bmatrix}
%	\hat{\mathcal{H}}_k(t)
%	\begin{bmatrix}
%		\opc{k,\Asmall} \\ \opc{k,\Bsmall}
%	\end{bmatrix}	\\
%\hat{\mathcal{H}}_k(t) &=  \R(k,t) \cdot \bpauli + (\n \cdot \bpauli) \otimes \hat{X}_k + \sum_l \hbar\omega_l \opbdag{k,l} \opb{k,l}
%\end{split}
%\end{align}
%
where $\n$ specifies a unit vector for the bath coupling, 
$\hat{X}_k = \sum_l \lambda_l (\opbdag{k,l} + \opb{k,l})$, $\lambda_l$ are coupling constants, and $[\opb{k,l},\opbdag{k',l'}]=\delta_{k,k'} \delta_{l,l'}$.
In terms of the original fermions, a $\pauli^z$-bath coupling, given by $\n=(001)$, 
would correspond to a term $(\opcdag{k,\Asmall}\opc{k,\Asmall} - \opcdag{k,\Bsmall}\opc{k,\Bsmall}) \otimes \hat{X}_k$.
This will be our standard choice unless otherwise specified.
Our dissipative Rice-Mele model can then be effectively regarded as a collection of $N$ independent dissipative 2-level systems, one 
for each momentum $k$ in the BZ, with a system-bath coupling effectively acting on $\pauli^z$. 
%with Hamiltonians:
%\begin{equation}	\label{Htot_k}
%	\Htotk(t) = \Hcalsysk(t) + \HcalSBk + \Hbathk
%\end{equation}
%where
%%, from Eq. \eqref{HSB_RMk}, it follows that 
%$\HcalSBk = \pauli_z \otimes \hat{X}_k$. 
The interaction between system and environment is encoded in the bath spectral function
${\mathcal J}(\omega) = \sum_l \lambda_l^2 \delta(\omega - \omega_l)$. 
We will consider a standard Ohmic dissipation~\cite{Leggett_RMP87}, modelled in the frequency continuum limit as 
${\mathcal J}(\omega) = 2 \alpha \hbar^2 \omega F_{\scriptscriptstyle\mathrm{cutoff}}(\omega,\omega_c)$,
where $\alpha$ is the (dimensionless) coupling strength, and $F_{\scriptscriptstyle\mathrm{cutoff}}(\omega,\omega_c)$ implements a suitable frequency cutoff $\omega_c$. 
In the following we will adopt either the exponential form $F_{\scriptscriptstyle\mathrm{cutoff}}=\exp( -\omega / \omega_c )$ 
or the hard-cutoff $F_{\scriptscriptstyle\mathrm{cutoff}}=\Theta(\omega_c-\omega)$.
%Here $\alpha$ is the coupling strength and $\omega_c$ is the cutoff frequency.

%----------------------------------------------
\section{The pumped charge}		\label{sec:charge}	
%----------------------------------------------
%%%%%%%%%%%%%%%%%%%%%%%%%%%%%%%%%%%%
%			OBSERVABLES
%%%%%%%%%%%%%%%%%%%%%%%%%%%%%%%%%%%%
%\subsection{The pumped charge}
%
The current density operator $\Jop(t)$ is obtained as a derivative of the system Hamiltonian 
with respect to an external flux $\Phi$ passing through the hole of the PBC ring, 
$\Jop = \frac{\partial \Hsys(\Phi)}{\partial \Phi} \vline_{\Phi = 0}$.
%$\Jop = \frac{1}{\hbar} \frac{\partial \Hsys(\kappa)}{\partial \kappa} \vline_{\kappa = 0}$, 
%where $\kappa = \frac{2\pi}{L}\frac{\Phi}{\Phi_0}$, with $L=Na$ system length and $\Phi_0$ the flux quantum. 
In the present case, the current density operator can be expressed as 
\begin{equation}	\label{H_RMdiag}
	\Jop(t) = \frac{1}{L} \sum_{k}^{\BZ}
	\begin{bmatrix}
		\opcdag{k,\Asmall} & \opcdag{k,\Bsmall}
	\end{bmatrix} 
	\, \Jkt_k(t) \,  
	\begin{bmatrix}
		\opc{k,\Asmall} \\ \opc{k,\Bsmall}
	\end{bmatrix}	
\end{equation}
%
%$\Jop(t) = \frac{1}{\hbar} \sum_{k}^{\\BsmallZ} \J(k,t) \cdot \bpauli$, 
where $L=Na$ is the system length,
\begin{equation}
\Jkt_k(t) = \frac{ea}{2\hbar} \bigg( J_2(t) \sin (ka) \pauli^x + \Big( J_1(t) - J_2(t) \cos (ka) \Big) \pauli^y \bigg) \;,
\end{equation}
and $e$ is the charge of the electron. 
Given the density matrix $\rhosysk(t)$, the pumped charge during the $m^{th}$ driving period, $Q_m$, is thus given by
integrating the current density over the period. In the thermodynamic limit $L\to\infty$ we can write
%
%\begin{equation} \label{RM_Qsingle}
%	Q_m = \frac{1}{2Na\hbar} \sum_k^{\BZ} \int_{(m-1)\period}^{m\period} dt \;
%	\Trace \Big\{  \Jkt_k(t) \, \rhosysk(t) \Big\} \;.
%\end{equation}
%
%\textcolor{red}{We are interested in the behaviour in the thermodynamic limit; here the sum over $k$ can be substituted by an integral in the form}
%
\begin{equation} \label{RM_Qsingle1}
	Q_m = \int_{-\frac{\pi}{a}}^{\frac{\pi}{a}} \frac{\ud k}{2\pi} 
	\int_{(m-1)\period}^{m\period} \ud t \; 	\Trace \Big( \Jkt_k(t) \, \rhosysk(t) \Big) \;.
\end{equation}
After a sufficiently large number of cycles, the average pumped charge is expected to converge to an asymptotic value
%value, which can be defined as: 
\begin{equation}	\label{RM_Qinfty}
	Q_m \stackrel{\scriptscriptstyle m\to \infty} {\longrightarrow}\Qinfty \equiv \lim_{M \rightarrow \infty} \frac{1}{M} \sum_{m=1}^M Q_m \;.
\end{equation}
In the next section we are going to use Floquet theory to study in detail how this convergence occurs.

%..................................................................................................................................................................%
\subsection{Floquet analysis for the pumped charge} \label{sec:floquet_coherent}
We introduce here some important notions of Floquet theory applied to charge pumping, extending the coherent evolution treatment presented
in Ref.~\cite{Privitera_PRL18}. 
Since the driving is periodic, \ie $\Hsys(t) = \Hsys(t+\period)$, from Floquet theory \cite{Shirley_PR65,Grifoni_PR98,Hausinger_PRA10} 
we know that the solutions to the time-dependent Schr\"odinger equation for the closed system have the following form
\begin{equation}	\label{Floquet}
	\ket{\Fstate_{n}(t)} = e^{-i \Fqe_{n} t/\hbar} \ket{\Fmode_{n}(t)}
\end{equation}
where ${n}$ labels the possible solutions, $\ket{\Fstate_{n}(t)}$ are called \textit{Floquet states}, 
$\ket{\Fmode_{n}(t)}$ are called \textit{Floquet modes} and are $\period-$periodic, and $\Fqe_{n}$ are the \textit{quasi-energies}. 
In the present case, using $k$ as a quantum number and $n=\pm$ for two Floquet states $\ket{\Fstate_{n}^{k}(t)}$ at each $k$, 
we can rewrite the density matrix $\rhosysk(t)$ in the coherent Floquet basis
\begin{equation}	\label{rhoFloquet}
	\rhosysk(t) = \sum_{n,{n'}} \rhoFlqt{n{n'}}(t) \, \ket{\Fstate_{n}^k(t)} \bra{\Fstate_{{n'}}^k(t)}
\end{equation}
where $\rhoFlqt{n{n'}}(t) = \bra{\Fstate_{n}^k(t)} \rhosysk(t) \ket{\Fstate_{{n'}}^k(t)}$. 

%
%Given the unitary evolution operator $U(t_2,t_1)$ induced by $\Hsys$, it follows that 
%$\ket{\Fstate_n(t_2)} = U(t_2,t_1) \ket{\Fstate_n(t_1)}$. 
In this framework, the infinite-time average pumped charge, Eq. \eqref{RM_Qinfty}, can be written as 
\begin{align}	\label{RM_QFloquet_gen}
	\Qinfty = \lim_{\scriptstyle M\to \infty} 
	\frac{1}{M}  \sum_{n,{n'}} \int_{-\frac{\pi}{a}}^{\frac{\pi}{a}}
	\frac{\ud k}{2\pi} 
	\int_0^{M \period} \hspace{-3mm} \ud t \;
	e^{-\frac{i}{\hbar}(\Fqe_{n}^k-\Fqe_{{n'}}^k)t} F_{n{n'}}^{k}(t)  
\end{align}
where
\begin{equation}
F_{n{n'}}^{k}(t) = \rhoFlqt{n{n'}}(t)  \, {\mathbb{J}}_{{n'} n}^{k}(t) 
\end{equation}
and
\begin{equation}
{\mathbb{J}}_{{n'} n}^{k}(t) = \bra{\Fmode_{{n'}}^k(t)} \Jkt_k(t) \ket{\Fmode_{{n}}^k(t)} 
\end{equation}
is the matrix element of the current operator between Floquet modes, hence an explicitly periodic quantity. 
In the coherent evolution case \cite{Privitera_PRL18}, the density matrix $\rhoFlqt{{n}{n'}}(t)$ turns out
to be {\em time-independent} and related to the initial state $\ket{\psi_k(0)}$ as 
\begin{equation}
  \rhoFlqtcoh{{n}{n'}} = \bra{\Fmode_{{n}}^k(0)} \psi_k(0) \rangle \langle \psi_k(0) \ket{\Fmode_{{n'}}^k(0)} \;.
\end{equation}
%
%If the evolution is \textit{non-dissipative}, notice that the coefficients $\rhoFlqt{{n}}{{n'}}$ are time-independent. 
In turn, if the quasi-energies are non degenerate, when ${n} \neq {n'}$, the $k$-integral will vanish in the limit $t\to\infty$, since the oscillating phase factors
$\nep^{-i(\Fqe_{{n}}^k-\Fqe_{{n'}}^k)t/\hbar}$ will lead to destructive interference cancellations.
More formally, this is a consequence of the Riemann-Lebesgue lemma applied to the $k$-integration, 
%of the regular function $\rhoFlqt{{n}{n'}}(t){\mathbb{J}}_{{n'}{n}}^{k}(t)$ times the oscillating factor, 
as explained in detail in Refs.~\cite{Russomanno_PRL12,Russomanno_2013}.

Using this result in Eq.~\eqref{RM_QFloquet_gen}, and exploiting the infinite-time integration, it follows that only the populations 
$\rhoFlqtcoh{{n}{n}}= |\bra{\Fmode_{{n}}^k(0)} \psi_k(0) \rangle|^2$ of the Floquet bands come into play, and
one arrives at the so-called \textit{Floquet diagonal ensemble}~\cite{Russomanno_PRL12}
\begin{equation}	\label{RM_Qdiag_coh}
	\Qinfty = \Qdiag = \int_{-\frac{\pi}{a}}^{\frac{\pi}{a}} 
	\frac{\ud k}{2\pi} \sum_{{n}} \rhoFlqtcoh{{n}{n}} 
	\int_0^\period \! \ud t \, {\mathbb{J}}_{{n}{n}}^{k}(t) \;.
\end{equation}
With a very similar application of the Riemann-Lebesgue lemma it is also possible to see that, in the thermodynamic limit, $Q_m$ defined in Eq.~\eqref{RM_Qsingle1} tends 
to $\Qdiag$ when $m\to\infty$.

In the dissipative case, $\rhoFlqt{{n}{n'}}(t)$ is generally time dependent. 
What we find --- and explicitly discuss in Sec.~\ref{sec:results} and Fig.~\ref{RM_FloquetDynamics} --- is that after a certain transient, 
because of dissipative effects, $\rhoFlqt{{n}{n'}}(t)$ becomes $\period$-periodic.  
Hence, we can again apply the Riemann-Lebesgue lemma as done in Ref.~\cite{Russomanno_2013} and show that only the diagonal terms contribute, arriving 
at the dissipative version of the Floquet diagonal ensemble formula for the average pumped charge
\begin{equation}	\label{RM_Qdiag_diss}
	Q_m \stackrel{\scriptscriptstyle m\to \infty} {\longrightarrow}\Qinfty = \Qdiagdiss =  
	\int_{-\frac{\pi}{a}}^{\frac{\pi}{a}} 
	\frac{\ud k}{2\pi} \sum_{{n}} 
	\int_0^\period \! \ud t \, \rhoFlqt{{n}{n}}(t) \, {\mathbb{J}}_{{n}{n}}^{k}(t) \,.
\end{equation}

As in Ref.~\cite{Privitera_PRL18}, the asymptotic pumped charge can be related to the properties of the Floquet quasienergies. 
In order to do that, a result coming from our numerics is crucial: if we approximate
$\rhoFlqt{{n}{n}}(t)$ at stationarity with its average value on one period $\overline{\rho}_{{n}{n}}^k$, we get corrections to the pumped charge of the order $\sim10^{-6}$. 
So, a very good approximation~\cite{note_RWA} consists in replacing $\rhoFlqt{{n}{n}}(t)$ with its time-average $\overline{\rho}_{{n}{n}}^k$. 
With this approximation, using arguments similar to those of Ref.~\cite{Privitera_PRL18}, we find that the asymptotic pumped charge can be written in the form
\begin{equation}	\label{RM_Qdiag_diss_floquet}
	\Qdiagdiss =  \frac{1}{\hbar\omega}\int_{-\frac{\pi}{a}}^{\frac{\pi}{a}} \ud k \, 
	\sum_{{n}} \, \overline{\rho}_{{n}{n}}^k \, \partial_k\Fqe_{{n}}^k \;,
\end{equation}
which looks very similar to Eq.~(4) in Ref.~\cite{Privitera_PRL18}, obtained for a coherent dynamics. 
Because the derivative with respect to $k$ can be recast as the derivative with respect to an external flux~\cite{Privitera_PRL18}, this equation is also strictly analogous to Eq.~(19) of Ref.~\cite{Russomanno_PRB11}, where pumping in a dissipative superconducting nanocircuit was considered.

%%%%%%%%%%%%%%%%%%%%%%%%%%%%%%%%%%%%
%	Comparison between methods
%%%%%%%%%%%%%%%%%%%%%%%%%%%%%%%%%%%%
\section{Numerical methods} \label{sec:methods}
In this section we introduce and compare two different methods for computing the dissipative dynamics of our model. 
The first method discussed in Sec.~\ref{QME:sec} is the standard Bloch-Redfield Quantum Master Equation (QME)~\cite{Cohen:book,Gaspard_JCP99a} for a two-level system, 
which is fast, but valid only within the weak-coupling Born-Markov approximation. 
The second method, described in Sec.~\ref{sec:ChainMapping}, is based on a unitary mapping of the harmonic bath into a bosonic chain interacting with the system
~\cite{JMath_Chin_2010,PRA_Vega_2015,PRB_Schroeder_2016,PRL_Tamascelli_2019}. 
The full system-plus-bath time evolution is then obtained by using Matrix Product States (MPS) and the 2-site Time-Dependent 
Variational Principle~\cite{PRL_Haegeman_2011,SIAM_Lubich_2015,PRB_Haegeman_2016}. 
To the best of our knowledge, we present here the very first results obtained by this method in a non-equilibrium scenario, \ie in a system with explicit time-dependence.
This second method is reliable at any coupling strength and accounts for non-Markovian effects, at the expense of a higher computational complexity. 
We therefore used it here just to benchmark the QME results in some specific regimes. 
We remark that the reliability of a QME was thoroughly tested in Ref.~\cite{Arceci_PRB17} against the numerically-exact QUasi-Adiabatic Path Integral (QUAPI) 
method~\cite{Makri_95, Makri_95_bis} for the dissipative Landau-Zener model.
%It proved to be very reliable at any temperature (even at zero temperature), if the weak-coupling condition was met. 
We perform here a similar check with the present driving scheme, see Sec.~\ref{sec:compare_QMEMPS}. 

All the results presented in Sec.~\ref{sec:results} have been obtained using the QME, since this allows us to check the thermodynamic limit ($N \to \infty$), 
in a reasonable computational time. 

%%%%%%%%%%%%%%%%%%%%%%%%%%%%%%%%%%%%
%				QME
%%%%%%%%%%%%%%%%%%%%%%%%%%%%%%%%%%%%
\subsection{Bloch-Redfield equation} \label{QME:sec}
According to the Hamiltonian in Eq.~\eqref{Htot_k} and under the assumptions of weak system-bath coupling and Born-Markov approximation
\cite{Cohen:book,Gaspard_JCP99a,Nalbach_PRA14,Javanbakht_15,Yamaguchi_PRE17,Arceci_PRB17}, 
we can write a QME to describe the reduced density matrix $\rhosysk(t)$ for the system for each $k$-vector.
There are many slightly different ways of writing down the relevant QME, depending, for instance, on whether or not one adopts a rotating wave approximation (RWA).
The approach used below makes use of a RWA and is essentially equivalent to the ``double-sided adiabatic QME'' in Lindblad form explained in Ref.~\cite{Zanardi_NJP12}.
%
% non-adiabatic QME by  Koslov, Rivas: not here!
Observe that this treats the dissipative dynamics assuming an adiabatic driving, but we will also apply it in regimes where this condition is slightly violated. 
%We might be losing the interplay between dissipation and non-adiabaticity but, 
%We expect this should not impact our results, at least qualitatively, whenever the driving is sufficiently slow.

Writing the system density matrix in the Bloch-vector notation 
$\rhosysk(t) = (\id + \bfr_k(t) \cdot \bpauli) / 2$,
the equations to solve for the dynamics in Schr\"odinger representation read
%can be split into two contributions, 
%respectively from coherent and incoherent time evolution: 
\begin{equation}	\label{EqsToSolve}
	\dot{\bfr} = \frac{2}{\hbar} \, \R \times \bfr  - \mathA_{\mathrm{diss}} \cdot \bfr - \bfb \;.
\end{equation}
where we dropped the $k$ and $t$ labels from $\bfr$ and related quantities.
From the Lindblad form derived using the rotating-wave-approximation (RWA) \cite{Zanardi_NJP12}, we can express the relevant ingredients
appearing in $\mathA_{\mathrm{diss}}$ and $\bfb$ in terms of pure-dephasing $\Gpuredeph$ and relaxation $\Grelax$ rates \cite{Schoen_PhyScr02}
\begin{subequations}
\begin{align}
	\label{PureDephasingRate}
	\Gpuredeph &= \frac{(\n \cdot \R)^2}{E^2} \frac{S_X(0)}{\hbar^2} \\
	\label{RelaxationRate}
	\Grelax &= \left(1-\frac{(\n \cdot \R)^2}{E^2}\right) \frac{S_X(2E/\hbar)}{\hbar^2}
\end{align}
\end{subequations}
where $S_X(\omega)=\gamma(\omega)+\gamma(-\omega)$ is the symmetrised Fourier transform of the free thermal bath correlation function
\begin{equation}
\gamma(\omega) = \int_{-\infty}^{+\infty} \ud t \; \nep^{i\omega t} \langle \hat{X}(t) \hat{X}(0) \rangle_0 \;.
\end{equation}  
For the case of Ohmic dissipation with exponential cutoff, ${\mathcal J}(\omega) = 2 \alpha \hbar^2 \omega \nep^{-\omega/\omega_c}$, we find that
$S_X(0) = 8\pi \hbar \alpha / \beta$ and $S_X(2E/\hbar) = 2\pi {\mathcal J}(2E/\hbar) \coth(\beta E)$, 
with $E = |\R|$, $\beta = (k_B \tmpr)^{-1}$ and $k_B$ the Boltzmann constant. 
In terms of these quantities, the dissipation matrix $\mathA_{\mathrm{diss}}$ and the vector $\bfb$ have the form
\begin{subequations}
\begin{align}
\mathA_{\mathrm{diss}} &= \left( 
	\begin{array}{lll} 	\GdephRWA{x} 		& \gammaRWA_{xy} 		& \gammaRWA_{xz}  \\
					\gammaRWA_{xy} 	& \GdephRWA{y} 		& \gammaRWA_{yz} \\
					\gammaRWA_{xz} 	& \gammaRWA_{yz} 		& \GdephRWA{z} 
	\end{array} \right) \\
	\bfb &= \frac{\R}{E} \Grelax \tanh(\beta E) \;,
\end{align}
\end{subequations}
where
\begin{subequations}
\begin{align}
	\GdephRWA{i} &=  \left( \frac{\Grelax}{2}+\Gpuredeph \right) + \frac{R_i^2}{E^2} \left( \frac{\Grelax}{2}-\Gpuredeph \right) \\
	\gammaRWA_{ij} &= \frac{R_i R_j}{E^2} \left( \frac{\Grelax}{2}-\Gpuredeph \right) \;.
\end{align}
\end{subequations}
These equations agree with those discussed in Ref.~\cite{Schoen_PhyScr02}, obtained, with similar approximations, 
from a perturbative diagrammatic approach. 
As one can easily verify, if $\R$ is time independent, the final steady state value of $\bfr(t\to\infty)$ correctly describes
the thermal density matrix for each momentum $k$. 
To compute the system's dynamics, we solved Eqs.~\eqref{EqsToSolve} through a standard fourth-order Runge-Kutta method. 

%%%%%%%%%%%%%%%%%%%%%%%%%%%%%%%%%%%%
%			Chain Mapping
%%%%%%%%%%%%%%%%%%%%%%%%%%%%%%%%%%%%
\subsection{Chain mapping based time evolution using Matrix Product States }
\label{sec:ChainMapping}
In order to benchmark the QME we employ a numerically exact technique using MPS, 
recently introduced by Tamascelli \textit{et al.}\,\cite{PRL_Tamascelli_2019}. The method is essentially based on two key ideas. First, a more efficient representation of the finite temperature bath is obtained by modifying the spectral density. 
In particular, the thermal state will be represented by the vacuum state of a transformed bosonic environment, which allows us to write the state of system-plus-bath as a 
pure state rather than a density matrix. 
Second, the spin-boson (star-like) Hamiltonian is transformed to a chain geometry, using orthogonal polynomials. 
This chain Hamiltonian can be simulated efficiently using MPS. 
For our benchmarks we use the Ohmic spectral density with hard cutoff: ${\mathcal J}(\omega) = 2 \alpha \hbar^2 \omega \Theta(\omega_c-\omega)$. 

Let us now discuss the method in more detail. 
We start from the continuum version of the Hamiltonian in Eq.~\eqref{Htot_k} for a given wave vector $k$. 
Parameterizing the modes by the dimensionless variable $x$ we have \cite{PRA_Vega_2015,PRB_Schroeder_2016}
\begin{equation} \label{H_k_continuous}
\hat{\mathcal{H}}_k(t) =  \R(k,t) \cdot \bpauli + (\n \cdot \bpauli) \otimes \int_{0}^{1} \ud x \ \lambda(x) (\opbdag{k}(x) + \opb{k}(x)) + \int_{0}^{1} \ud x \ \hbar \omega(x) \opbdag{k}(x) \opb{k}(x)
\end{equation}
%
%\RED{I think we should use the previous notation as much as possible, not to confuse the reader. Maybe we should write explicitly how $x$ and $l$ are related, and use 
%$\lambda(x)$ and $\omega(x)$. What do you think? } 
where now the bosonic operators have a (frequency) continuum normalisation in terms of Dirac delta: $[ \opb{k}(x), \opbdag{k'}(x') ] =\delta_{k,k'} \delta(x-x')$. 
Here, the functions $\lambda(x)$ and $\omega(x)$ define the interaction and dispersion, respectively, and are connected to the spectral density through the relation 
$\lambda^2(x)=\omega'(x) {\mathcal J}(\omega(x))$~\cite{JMath_Chin_2010,PRB_Schroeder_2016}. 
The freedom in choosing those functions is exploited by setting $\omega(x)=\omega_{c} x$, which is useful in the algebra of orthogonal polynomials performed below.  
At this point the bath is described by the density matrix corresponding to the thermal state, requiring large local dimensions at high temperatures. 
Now the new insight of Ref.~\cite{PRL_Tamascelli_2019} --- see also Ref.~\cite{PRA_Vega_2015} for an alternative route through a thermal Bogoliubov transformation --- 
is that the thermal state can be represented by the bosonic vacuum, if the spectral density 
${\mathcal J}(\omega)$ is replaced by the temperature-dependent spectral density
\begin{equation}
{\mathcal J}_{\beta}(\omega) = \frac{\mathrm{sgn}(\omega){\mathcal J}(|\omega|)}{2}\left( 1+ \coth(\beta\omega/2) \right) \;,
\end{equation}
extending its support to negative frequencies. 
As a consequence, the integrals in Eq.~\eqref{H_k_continuous} have an extended domain $[-1,1]$, and $\lambda(x)$ has to be replaced by 
$\lambda_\beta(x)$, which is defined through the thermal spectral density: $\lambda_{\beta}^2(x)=\omega_c {\mathcal J}_{\beta}(\omega(x))$.
%${\mathcal J}_{\beta}(\omega)$ and $\omega(x)$ as before. 
%Alternatively this result can be obtained by performing a thermal Bogoliubov transformation (see Ref.~\cite{PRA_Vega_2015}) and 
%recombining the resulting two baths before performing the chain mapping. Note that at zero temperature the thermal spectral density reduces to the original one.

Next, following Refs.~\cite{PRL_Prior_2010,JMath_Chin_2010,PRB_Schroeder_2016}, we perform a mapping from the star-like spin-boson geometry to a chain geometry. 
We do that by defining the unitary transformation
\begin{equation} \label{chain_transform}
\opadag{k,j} = \int_{-1}^{1}\ud x \, U_{j}(x)\opbdag{k}(x)	 \quad \text{with} \quad U_{j}(x)=\lambda_{\beta}(x)p_j(x)
\end{equation}
and inverse transformation $\opbdag{k}(x)=\sum_{j=0}^{\infty} U_{j}(x)\opadag{k,j}$. 
The (real) polynomials $p_j(x)$ of degree $j$ are orthonormal with respect to the inner product 
\begin{equation}
\langle p_j , p_{j'} \rangle = \int_{-1}^{1} \ud x \ \lambda_{\beta}^2(x) \, p_j(x) \, p_{j'}(x) = \delta_{j,j'} \;. 
\end{equation}
Moreover, the corresponding monic polynomials $\pi_{j}(x)$ --- obtained when dividing each polynomial $p_j$ by the coefficient of the leading power --- 
satisfy the three term recurrence relation \cite{JMath_Chin_2010}
\begin{equation}
\pi_{j+1}(x) = (x-\alpha_{j})\pi_{j}(x)-\beta_{j}\pi_{j-1}(x) \qquad j=0,1,2,...
\end{equation}
with initial polynomial $\pi_{-1}\equiv 0$. While those polynomials are not needed explicitly, we need to find the recurrence coefficients $\alpha_{j}$ and $\beta_{j}$. 
For zero temperature they are known for some special spectral densities \cite{JMath_Chin_2010}, but in most cases we need to calculate them numerically~\cite{ACM_Gautschi_1994,Gautschi_2004}. Exploiting this recurrence relation, the transformation in Eq.~\eqref{chain_transform} yields the chain geometry Hamiltonian
\begin{equation}
\hat{\mathcal{H}}_k(t) =  \R(k,t) \cdot \bpauli + (\n \cdot \bpauli) \otimes \kappa_0 (\opadag{k,0} + \opa{k,0}) + 
\sum_{j=1}^{\infty} \kappa_j(\opadag{k,j}\opa{k,j-1} + \hc) + \sum_{j=0}^{\infty}\hbar\Omega_j \opadag{k,j} \opa{k,j} 
\end{equation}
with system-bath coupling $\kappa_0=\sqrt{\int_{-\omega_c}^{\omega_c} \ud \omega \ {\mathcal J}_{\beta}(\omega)}$, hopping amplitudes 
$\kappa_j=\hbar\omega_c\sqrt{\beta_{j}}$ and on-site energies $\Omega_j=\omega_c\alpha_{j}$. 
This infinite chain can be truncated at some finite length $L_{\mathrm{chain}}$, such that the excitations reflected at the border of the truncated chain 
do not reach the system~\cite{PRB_Schroeder_2016}.
We then carry out the time evolution using the 2-site Time Dependent Variational Principle~\cite{PRL_Haegeman_2011,SIAM_Lubich_2015,PRB_Haegeman_2016,}, 
which has proven to combine high accuracy and efficiency~\cite{Arxiv_Schollwoeck_2019}. Let us now briefly discuss the sources of errors in this method. First, we need to find the recurrence coefficients for the chain mapping, which is done numerically using the package in Refs.\cite{ACM_Gautschi_1994,Gautschi_2004}. The error, however, is under full control and any target accuracy can be reached. Once we have our chain Hamiltonian, we carry out time-evolution. Here, additional errors are due to the splitting of the differential equation in the Time-Dependent Variational Principle and due to limited representation capabilities of the MPS. While the splitting error can be controlled through the time step, the representation capabilities of the MPS are enhanced by increasing the bond dimension and the local bosonic dimension, which has to be truncated to some finite value. 
In summary, all errors are well under control and can be made arbitrarily small.

%%%%%%%%%%%%%%%%%%%%%%%%%%%%%%%%%%%%
%  Comparison between QME and MPS
%%%%%%%%%%%%%%%%%%%%%%%%%%%%%%%%%%%%
\subsection{Benchmarking the QME} \label{sec:compare_QMEMPS}
\begin{figure}
	\centering
	\includegraphics[width=\columnwidth]{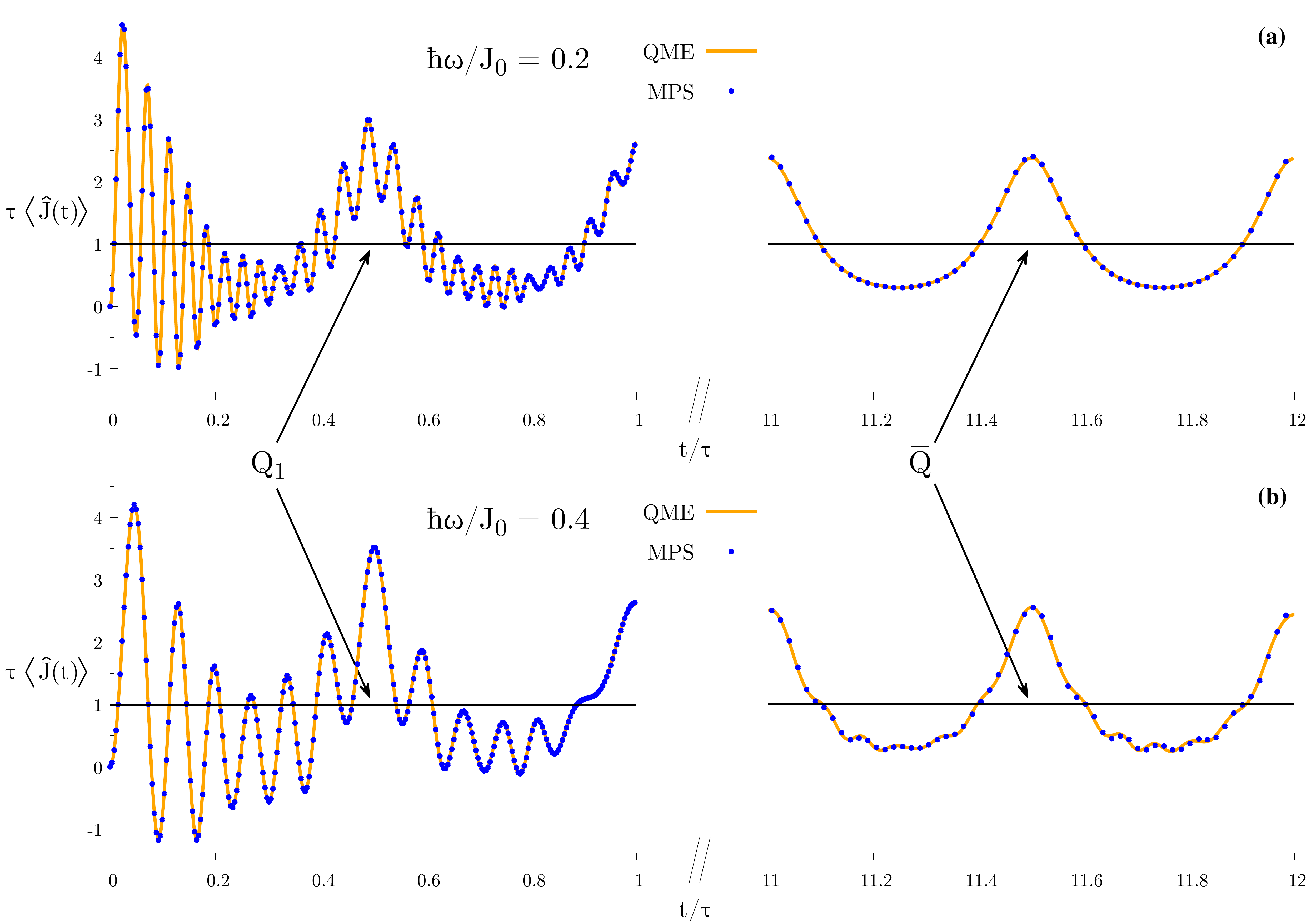} 
	\caption{
	Comparison of the dynamics of the total current computed through the QME approach (lines) and the MPS technique (dots) for 
	$\hbar \omega = 0.2 J_0$ (a) and $\hbar \omega = 0.4 J_0$ (b). 
	Data on the left (right) correspond to the first (last) period of pumping. 
     	The results refer to a chain with $N = 16$ diatomic cells and $M = 12$ pumping cycles, 
	with bath parameters $\tmpr = 0$, $\alpha = 0.001$ and $\hbar\omega_c = 10 J_0$ (for a spectral function with hard cutoff, see Sec.~\ref{sec:ChainMapping}). 
	We observe an excellent agreement between the two approaches. 
	For the MPS calculations we used a bond dimension $D=10$, a local bosonic dimension $d=4$ (at most 3 bosons per site), and a bath harmonic chain of length 
	$L_{\mathrm{chain}}=1030$ (for $\hbar\omega=0.2J_0$) or $L_{\mathrm{chain}}=600$ (for $\hbar\omega=0.4J_0$). Convergence was observed for these parameters.
}
\label{Fig_MPSQME}
\end{figure}
%
%\RED{New part written by Luca}
%
We benchmark here the QME data against the numerically-exact MPS approach. 
Since the latter technique is computationally heavy, we consider here a chain of $N=16$ diatomic cells. 
Moreover, the most interesting results, in the following, will be related to a low-temperature bath, which is also the regime were the
QME is believed to be less reliable~\cite{Weiss:book}. For this reason, we fix here $\tmpr=0$. 
The bath spectral function is again Ohmic with hard cutoff, as described in Sec.~\ref{sec:ChainMapping}, with $\alpha=0.001$ and $\hbar\omega_c = 10 J_0$. 
The pumping is performed over $M=12$ cycles and for two different driving frequencies $\hbar\omega/J_0 = 0.2$ and  $\hbar\omega/J_0 = 0.4$. 
In Fig.~\ref{Fig_MPSQME} we plot the total current density 
%properly rescaled by $\period$ --- 
as a function of time, focusing on the first and the last period, where stationarity is already attained. 
We observe an excellent agreement between the QME data (orange lines) and the MPS results (blue circles) in all regimes. 
The corresponding pumped charges agree within an error of the order of the precision we use to compute the $k$-integrals. 
Indeed, for $\hbar\omega /J_0 = 0.2$ we obtain $\Qinfty = 0.99999$ from the QME and $\Qinfty = 0.99998$ from the MPS, 
while for $\hbar\omega /J_0 = 0.4$ we get $\Qinfty = 1.00066$ from the QME and $\Qinfty = 1.00068$ from the MPS. 
%
%These results show a high reliability of our QME in predicting the correct system's dissipative dynamics, even at $\tmpr=0$. 
%We will therefore use it in the next sections to carry out a detailed analysis of the dissipative quantum pumping in our model. 

%-------------------------------------------------
\section{Dissipative pumping results} \label{sec:results}
%-------------------------------------------------
%
\begin{figure}
\centering
\begin{tabular}{c}
\includegraphics[width=80mm]{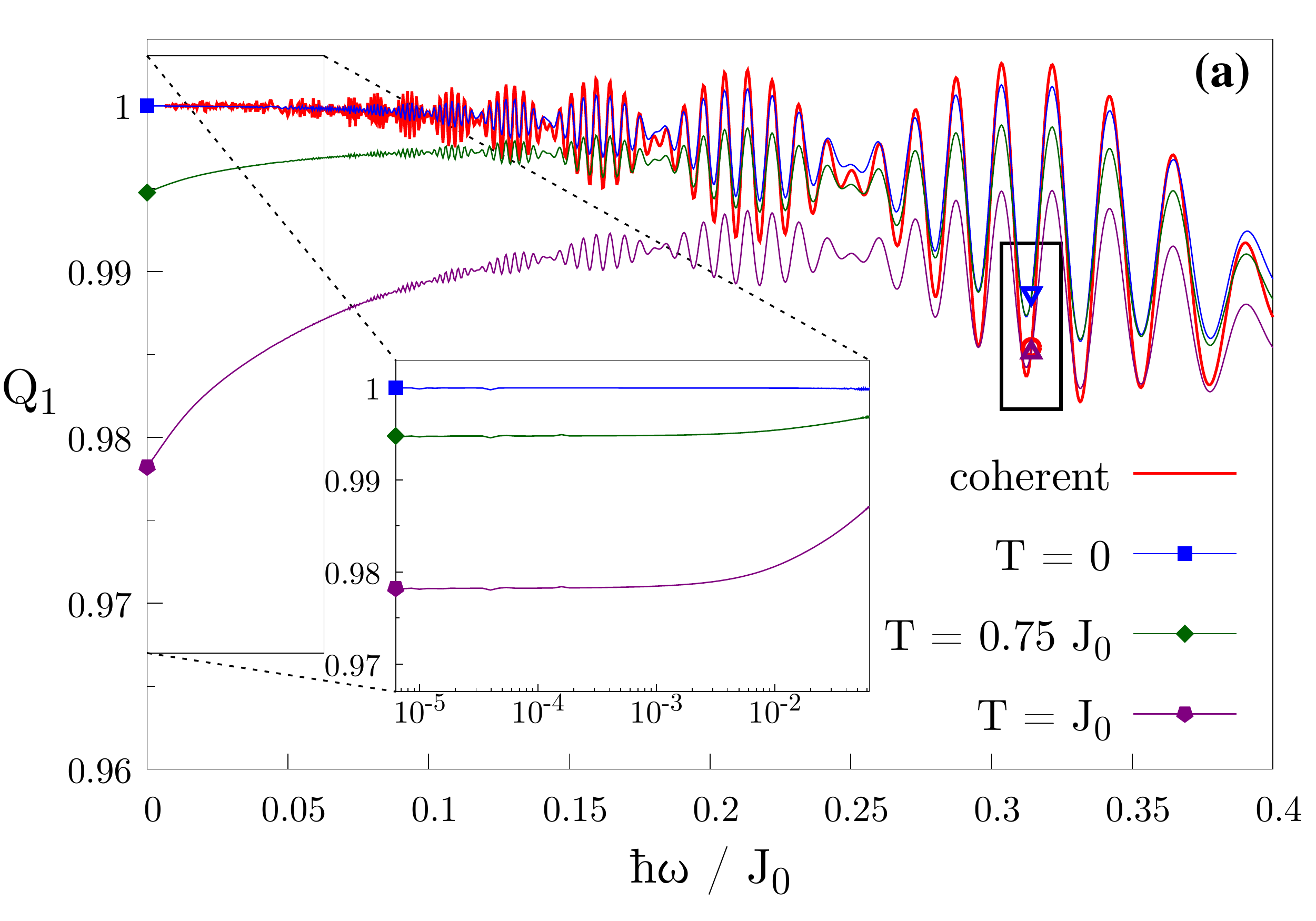}\\
\includegraphics[width=80mm]{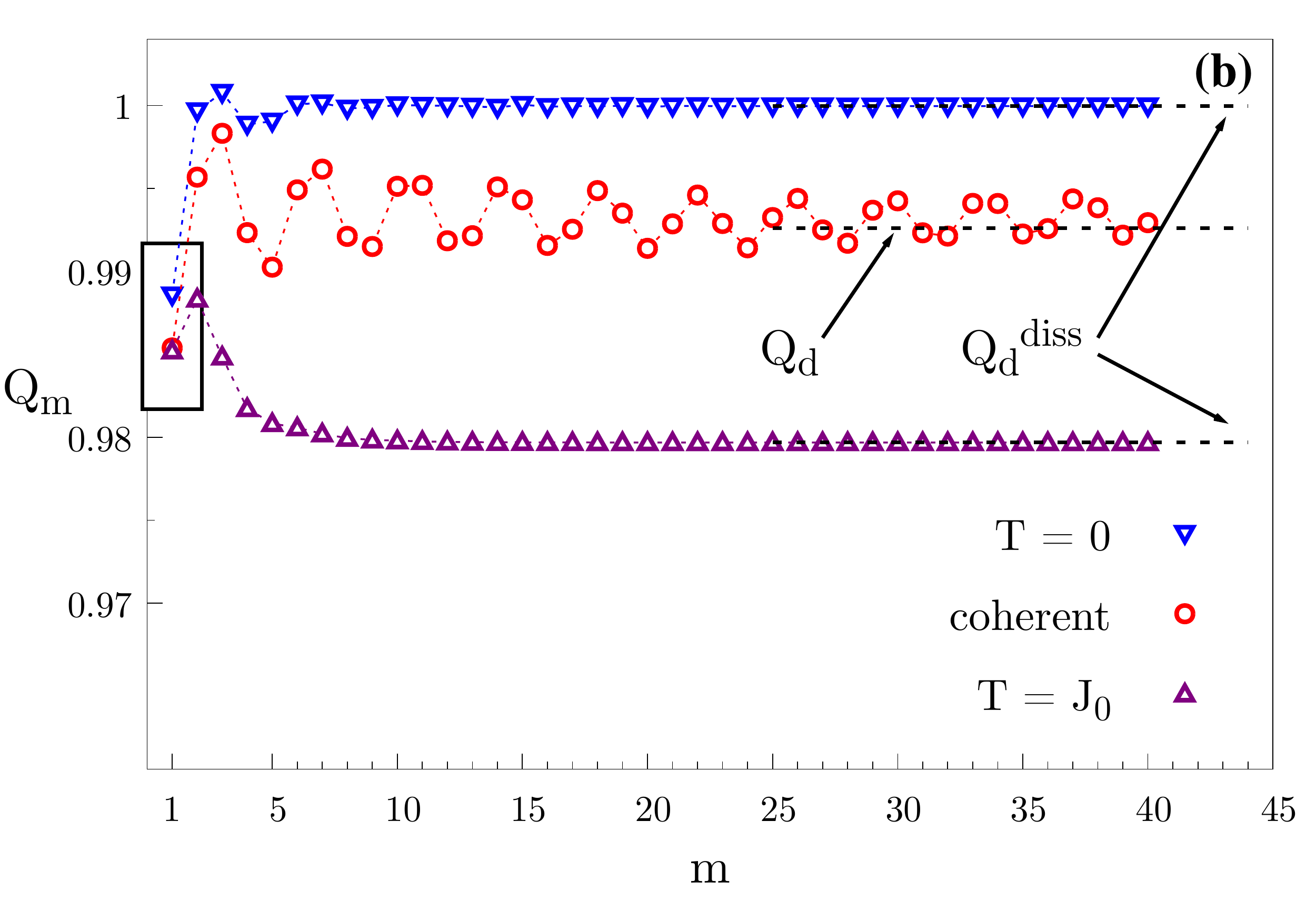}\\
\includegraphics[width=80mm]{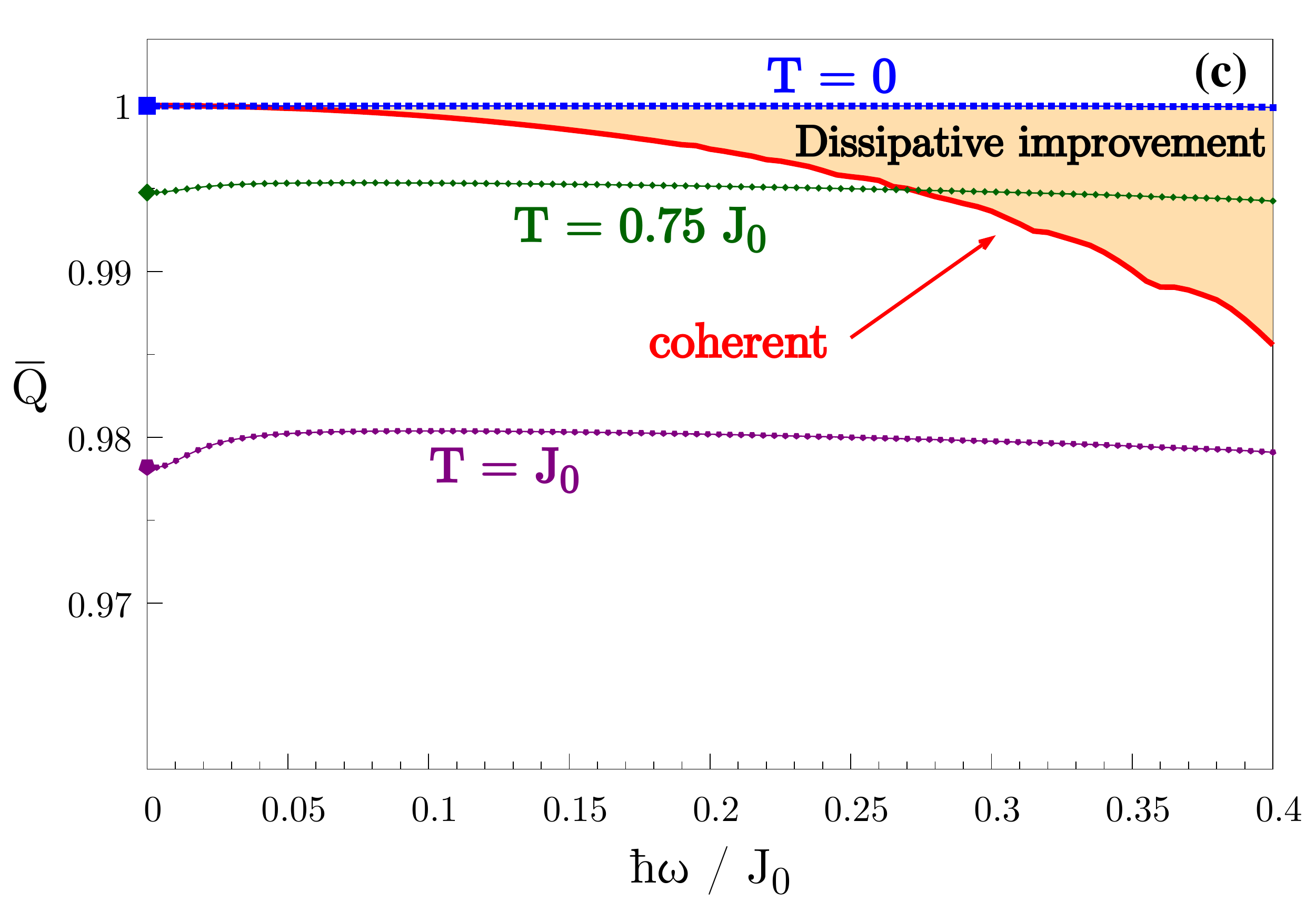}
\end{tabular}
\caption{
\textbf{\textrm{ (a)}} Pumped charge over the $1^{st}$ period $Q_1$ \vs driving frequency $\hbar\omega/J_0$, at bath temperatures $\tmpr$ ranging from $0$ to $J_0$,
compared to the coherent-evolution results of Ref.~\cite{Privitera_PRL18}.  
{\it Inset}: zoom of the $\omega\to 0$ region, showing the convergence to finite values depending on $\tmpr$.
\textbf{\textrm{ (b)}} Charge pumped in the $m^{th}$ period $Q_m$ \vs the cycle number $m$ for the coherent case (circles) compared to two dissipative evolutions (triangles)
at different bath temperatures $\tmpr$. 
Here $\period = 20\hbar/J_0$, as for the $m=1$ results in the rectangle shown in (a).  
The three horizontal dashed lines are the corresponding values from the Floquet diagonal ensemble, 
Eq.~\eqref{RM_Qdiag_coh} or Eq.~\eqref{RM_Qdiag_diss}. 
Notice that oscillations in the dissipative evolutions are damped much more rapidly. 
\textbf{\textrm{ (c)}} Average pumped charge in the infinite-time limit $\Qinfty$ \vs driving frequency $\hbar\omega/J_0$, at bath temperatures $\tmpr$ ranging from $0$ to $J_0$, 
compared to the Floquet diagonal ensemble value for the coherent evolution $\Qdiag$ in Eq. \eqref{RM_Qdiag_coh}. 
The region of dissipation assisted improvement over the closed-system pumping is highlighted by a yellow background.
}
\label{Figure1}
\end{figure}
We present here how dissipation affects the pumped charge at different driving frequencies. 
In all the following numerical analysis we will approximate the integral over $k$ with a discrete sum in the first BZ.
All the calculations are performed with sizes $N$ which we have verified to be large enough to be representative of the thermodynamic limit:
in practice, $N\sim 100$ is enough in the presence of dissipation.   
We choose the cutoff frequency $\omega_c$ in the spectral function 
to be much bigger than the widest spectral gap, $\omega_c = 1000 J_0/\hbar$ (we comment upon different choices of $\omega_c$ in \ref{AppRM_alpha}).
We start considering the bath coupling strength to be $\alpha = 0.001$; as we will see in Sec.~\ref{coupling:sec}, increasing it will have quite interesting effects on the pumping dynamics, stabilizing the
quantised pumping against the driving for $\hbar \omega > 0.4 J_0$. On the opposite, for smaller frequencies, results are insensitive on $\alpha$.
%We observe that the stationary pumped charge $\Qinfty$ converges, as $\alpha\to 0$, towards a well-defined limiting value, see \ref{AppRM_alpha} and Fig.~\ref{AppRM_AlphaScaling}.
%Our choice of interaction strength aims at capturing this limit, coherently with the weak-coupling regime in which our approach is valid.

Let us start considering the behaviour of the pumped charge after a single cycle. 
In Fig.~\ref{Figure1}(a) we plot the charge pumped after a single cycle, $Q_1$, versus the driving frequency $\omega$. 
On the one hand,
at larger values of the frequency, the bath has almost no effect, and the behaviour at all temperatures remains almost identical to the coherent one, 
which coincides with that reported in Ref.~\cite{Privitera_PRL18}. 
On the other hand, at smaller values of the frequency, the system has enough time to ``feel'' thermal effects and in general moves away from the ideal quantised pumping, 
here corresponding to $Q=1$. The charge converges to a finite value which depends on the bath temperature $T$, see the inset of Fig.~\ref{Figure1}(a).
We will further comment on this point later.
%These converged values are related to the fact that, for an extremely long period, the system has enough time to reach the stationary state within one cycle. 

These results change remarkably when pumping over a larger number of cycles. We find, and this is one of the main results of the paper, 
that the charge pumped over the $m$-th cycle $Q_m$ can overcome the corresponding coherent result in presence of a thermal bath 
of sufficiently low temperature. This is shown in Fig.~\ref{Figure1}(b).
%may become higher than the coherent evolution result.
%
%Indeed, Fig.~\ref{Figure1}(b) shows the pumped charge at the $m^{th}$ cycle as a function of the cycle number $m$. 
Observe that dissipation makes the convergence to the infinite-time average much faster than the coherent case. 
Notice also that the infinite-time average results are precisely described by the Floquet diagonal ensemble formulas, 
Eqs.~\eqref{RM_Qdiag_coh} and \eqref{RM_Qdiag_diss}, shown by horizontal dashed lines. 

Figure~\ref{Figure1}(c) shows the infinite-time pumped charge $\Qinfty$ versus the driving frequency $\omega$, for both coherent and dissipative evolutions
at different $\tmpr$.   
The dissipative results are obtained either from Eq.~\eqref{RM_Qinfty} (with $M=100$) or equivalently from Eq.~\eqref{RM_Qdiag_diss}; 
the coherent results are obtained from Eq.~\eqref{RM_Qdiag_coh}. 
Observe that at $\tmpr=0$ the dissipative results are always well above the coherent ones. 
We remark that dissipation at $\tmpr=0$ restores a nearly quantised pumped charge $\Qinfty=1$ {\em away} from the strict adiabatic limit $\omega\to 0$. 
$\Qinfty$ starts to deviate from one only for frequencies around $\hbar\omega \simeq 0.5 J_0$, as discussed in Sec.~\ref{coupling:sec}. 
This dissipative improvement of the pumped charge persists also at finite $\tmpr$, for large enough $\omega$.
We define this phenomenon \textit{dissipation assisted Thouless pumping}. 
This finding is supported by the benchmark done in Sec.~\ref{sec:compare_QMEMPS}, where the QME and the MPS results are found to be in excellent agreement.
% give an agreement of order $10^{-5}$ on the pumped charge, whose deviations from the quantised value are only of order $10^{-5}$. 
%
Our finding is also {\em qualitatively} independent of whether one assumes in the QME some form of RWA or not, as shown in~\ref{App_Eqs}. 
%where we compare different forms of RWA and the results obtained by QME without RWA. 
Notice, however, that some of the {\em quantitative} aspects of our results {\em do depend} on the details of the QME chosen.  
%
%\RED{Within the weak-coupling Born-Markov scheme, }
%This finding is independent of further approximations used in deriving the QME, \RED{as shown in}~\ref{App_Eqs}, where we compare different forms of RWA and the case without RWA, seeing results in good qualitative agreement (although not quantitative). 

Interestingly, in the small frequency regime we observe that $\Qinfty(\omega \rightarrow 0) \equiv Q_m(\omega \rightarrow 0)$ for any $m \geq 1$, 
\ie the $\omega\to 0$ limit is independent of the number of driving periods. 
This is because in the $\omega\to 0$ limit the dissipative transient induced by the bath occurs within a single driving period. 

\begin{figure}
\centering
	\includegraphics[width=.7\columnwidth]{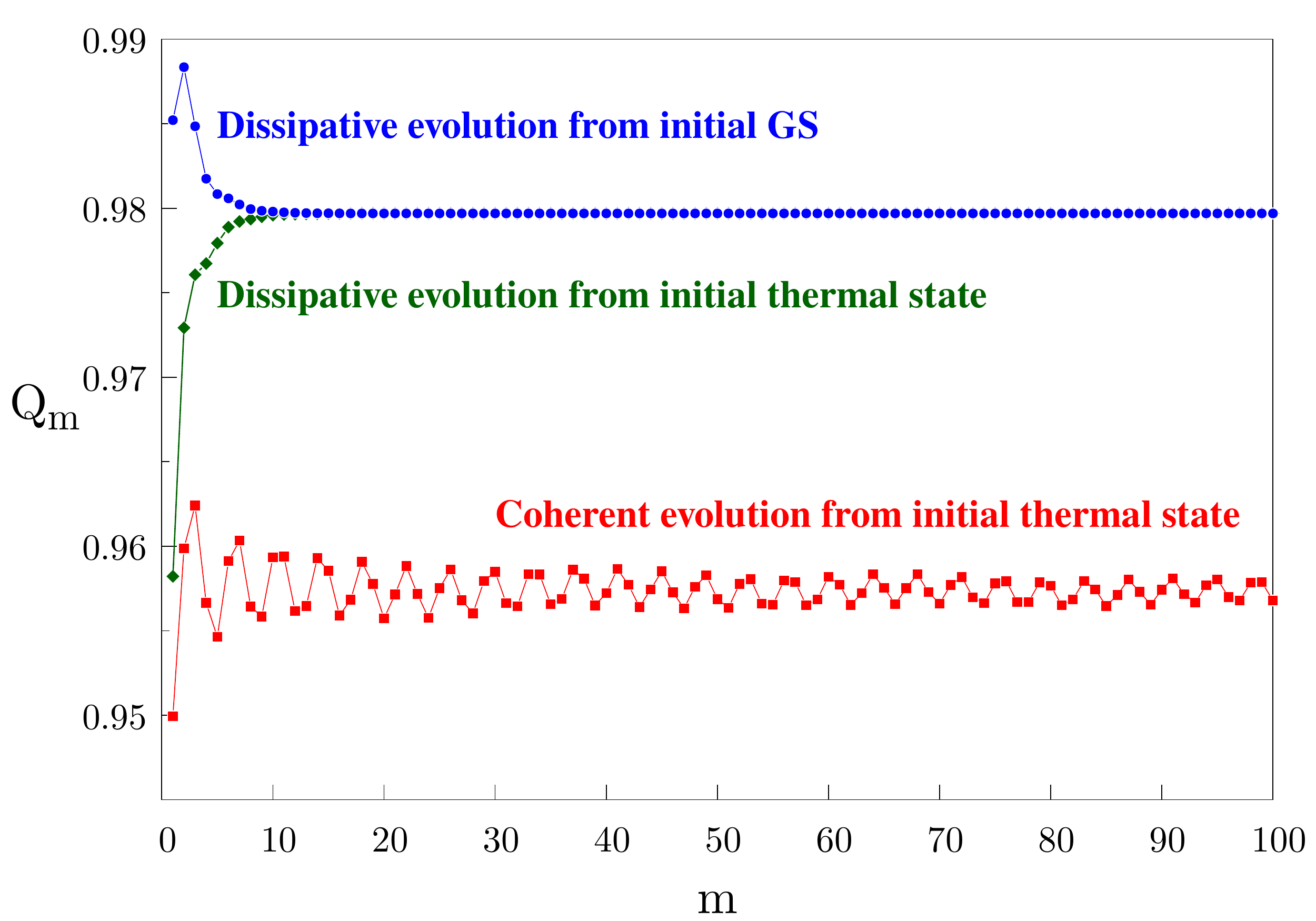}
	\caption{
	Comparison between results from the coherent evolution starting from an initial thermal state at $\tmpr = J_0$ (red squares) and the dissipative dynamics 
	induced by a bath at $\tmpr = J_0$, both starting from an initial ground state (blue circles) and from the thermal state at the same bath temperature (green diamonds). 
	Here we have a driving period $\period = 20 \hbar/J_0$. 
	The dissipative and coherent results are completely different. 
	Notice that the two dissipative evolutions converge to the same stationary state.
	}
	\label{RM_ClosedThermal_cycles}
\end{figure}

Notice that previous discussions of thermal effects in the Rice-Mele model~\cite{Wang_PRL13,Privitera_PRL18,Bardyn_PRX18} have considered a coherent Schr\"odinger
dynamics starting from an initial thermal state. 
We compared the results obtained from such an approach with the genuinely dissipative evolution studied in the present paper.
In general, we observed completely different results, both in the short and in the intermediate frequency ranges of study, 
as illustrated in Fig.~\ref{RM_ClosedThermal_cycles}.
This is definitely not surprising, but worth mentioning. 
Fig.~\ref{RM_ClosedThermal_cycles} is important also in another respect: we see that in the dissipative case, if we take very different initial conditions, we get the same asymptotic regime. 
This is not at all surprising in a dissipative system and marks the difference with the asymptotic regime of the coherent case~\cite{Russomanno_PRL12}.

\begin{figure}
\centering
	\includegraphics[width=\columnwidth]{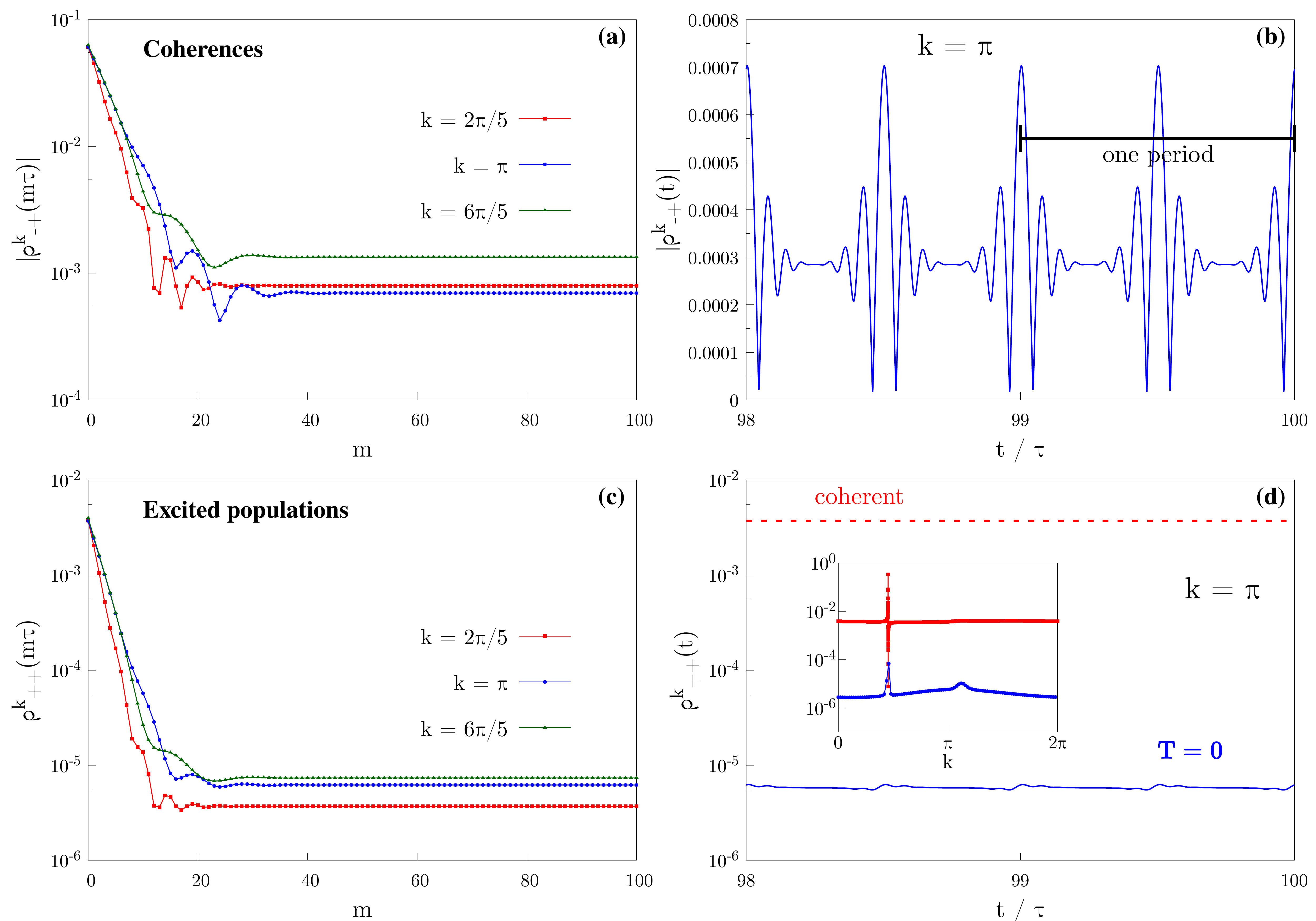}
	\caption{\textbf{Left panels}:
	Stroboscopic dynamics of: 
	\textbf{(a)} the coherences absolute value $|\rhoFlqt{-+}(m\tau)|$ and \textbf{(c)} the excited-Floquet band populations $\rhoFlqt{++}(m\tau)$, 
	for $k = \{2/5, 1, 6/5\} \pi$. Here $\tmpr = 0$ and $\period = 20 \hbar/J_0$. 
	%Notice the convergence to constant values after an adequately high number of cycles $m$. 
	%The phase of the off-diagonal terms tends also to a constant (not shown). 
	\textbf{Right panels}:
	Intra-period dynamics of: \textbf{(b)} $|\rhoFlqt{-+}(t)|$ and \textbf{(d)} $\rhoFlqt{++}(t)$ after stationarity is reached, for $k=\pi$. 
	All the other parameters are fixed as in panels (a) and (c). 
	Observe the $\period/2$-periodicity.
	In (d), the red horizontal dashed line shows $\rhoFlqt{++}$ for the corresponding coherent evolution, much higher than the dissipative result at $\tmpr=0$.  
	Inset in \textbf{(d)}: Comparison between $\rhoFlqt{++}$ for the coherent case (red squares) and the time-average $\overline{\rho}_{+ +}^k$ for the dissipative case 
	(blue circles) versus $k$.  
	%For all momenta, the population of the excited Floquet band in the dissipative case is some orders of magnitude smaller than the one for the coherent case.
	}
	\label{RM_FloquetDynamics}
\end{figure}

Focusing again on Fig.~\ref{Figure1}, observe the flat $\omega$-behaviour of $\Qinfty$ for the dissipative evolution at $\tmpr=0$. It would be interesting to pin down if the corrections to 
the strict adiabatic limit $\Qinfty(\omega\to 0)=1$ change from power-law \cite{Privitera_PRL18} to exponentially small in the presence of zero-temperature dissipation. 
Unfortunately, this question is extremely difficult to answer numerically.
Indeed, this aspect of the analysis is highly sensitive to the type of weak-coupling approximation performed in the QME. 
Although the QME results obtained with different flavours of RWA and without RWA are in good qualitative agreement, they are quantitatively different in that respect (see~\ref{App_Eqs}). 
A similar question might be asked concerning the behaviour of $\Qinfty(\omega\to 0, \tmpr)$ as a function of the bath coupling temperature $\tmpr$, a question 
that is once again numerically elusive and rather sensitive to the details of the QME used. 
A detailed MPS analysis might help in clarifying these points, but we suspect that the high numerical accuracy necessary to answer some of these questions would  
make these issues extremely delicate.  

%-........................................................................................................................................................%
\subsection{Floquet analysis of dissipative results}
It is insightful to understand the dissipative improvement shown in Fig.~\ref{Figure1}(c) within a Floquet framework. 
Let us therefore go back to the discussion of Sec.~\ref{sec:floquet_coherent} and study the dynamics of the system density matrix in the Floquet basis.

A crucial point will be to numerically show that the the populations and the coherences in the Floquet basis $\rhoFlqt{nn'}(m\tau)$ converge to an asymptotic $\tau$-periodic behaviour, in order to justify the application of the Riemann-Lebesgue lemma and the validity of Eq.~\eqref{RM_Qdiag_diss}.

Another point will be to understand the dissipation assisted Thouless pumping in terms of these quantities. In order to do that, we consider the lowest-energy Floquet state. This state is the Floquet state with maximum overlap with the instantaneous ground state. The instantaneous ground state gives rise to perfectly quantised pumping in the full adiabatic limit $\omega\to 0$. For finite $\omega$ the lowest-energy Floquet state does the same (up to corrections exponentially small in $\omega$)~\cite{Shih_PRB94,Privitera_PRL18}. We can construct the 
lowest-energy Floquet state by choosing, for each $k$, the Floquet state with (period-averaged) lowest-energy expectation $\ket{\psi_-^k(t)}$.  
On the opposite, choosing for each $k$ the Floquet state with period-averaged highest energy expectation we construct the highest energy Floquet state $\ket{\psi_+^k(t)}$. 
%The lowest-energy Floquet state is important because in the unitary case it is the state which pumps the charge nearest to the adiabatic quantised value.

It is interesting to understand if the dissipative improvement of the pumped charge is induced by the environment making this special state more populated. In order to answer to this question we will focus on the population of the highest-energy Floquet state $\rhoFlqt{++}(m\tau)$ and show that it is reduced by the environment in comparison to the unitary case. This means that the population of the lowest-energy state $\rhoFlqt{--}(m\tau)=1-\rhoFlqt{++}(m\tau)$ is increased.

We analyze the two points raised above in
Fig.~\ref{RM_FloquetDynamics}. We start focusing on the point about convergence. We show the stroboscopic dynamics of the coherences absolute values $|\rhoFlqt{-+}(m\tau)|$ between the two Floquet bands in Fig.~\ref{RM_FloquetDynamics}(a) and of the excited-band populations 
$\rhoFlqt{++}(m\tau)$ in Fig.~\ref{RM_FloquetDynamics}(c). We see that both quantities converge to stationary values for $m\to \infty$.  
The phases of the coherences (not shown) also converge to fixed values. 
After stroboscopic stationarity is reached, the intra-period behaviour of these quantities is illustrated in panels (b) and (d), respectively:  
observe a $\period/2$-periodicity in both cases.
%Summarizing, the system attains a $\period/2$-periodic steady state after a transient. 
This justifies the application of the Riemann-Lebesgue lemma in Sec.~\ref{sec:floquet_coherent} and the validity of Eq.~\eqref{RM_Qdiag_diss}. 
%Incidentally, from panel (d) we observe that the intra-period oscillations of $\rhoFlqt{++}(t)$ at stationarity are very small, so that we can actually approximate the 
%function with its mean over one period, as done in Eq. \eqref{RM_Qdiag_timeaverage}. 

%These findings justify the derivation of Eq. \eqref{RM_Qdiag_diss} in Sec. \ref{sec:floquet_coherent} for computing the stationary pumped charge 
%through a dissipative diagonal ensemble. 

Now let us focus on the point about the population of the highest-energy Floquet state.
In Fig.~\ref{RM_FloquetDynamics} (d) we compare the populations of the highest-energy 
Floquet band in the coherent and dissipative cases, showing that $\rhoFlqt{++}$ is generally reduced by several orders of magnitude 
in presence of dissipation at $\tmpr=0$. Hence $\rhoFlqt{--}$, the population of the lowest-energy Floquet state, is increased and then the topological pumping is improved at finite frequencies. 
In the main figure we fix $k$ and look at the dependence on time, while in the inset we plot the period-averaged $k^{th}$-population 
$\overline{\rho}_{++}^k$ versus the momentum $k$.
In the inset we note, incidentally, the presence of a value of $k$ where the coherent value shows an irregularity and the dissipative value shows a peak. 
That peak corresponds to a Floquet quasi-resonance which gives rise to a non-adiabaticity and increases the asymptotic dissipative population of the highest-energy 
Floquet band~\cite{Russomanno_2017}.

We conclude that dissipation moves the system towards the lowest-energy Floquet state, and that is the way in which the bath improves the topological pumping. This asymptotic condition where the Floquet state with maximum overlap with the instantaneous ground state is populated is very different from what happens in the high-frequency limit where the asymptotic condition is given by the Floquet-Gibbs states~\cite{Shirai_2015,Shirai_2016}.
%\ie the state built by taking all the Floquet states that correspond to the Floquet modes forming the lowest-energy Floquet band. 
%This state is the one closest to the adiabatic ground state, and 
%is the one which pumps a charge equal to the topological value, up to corrections exponentially small in the frequency~\cite{Privitera_PRL18,Shih_PRB94}. 
%In some sense, dissipation selects the Floquet state with lowest energy and topological pumping; in this way the pumped charge gets closer to the topological value. %compared to the coherent case.

\subsection{Effect of the bath coupling strength} \label{coupling:sec}
\begin{figure}
	\centering
        \begin{tabular}{c}
	  \includegraphics[width=9cm]{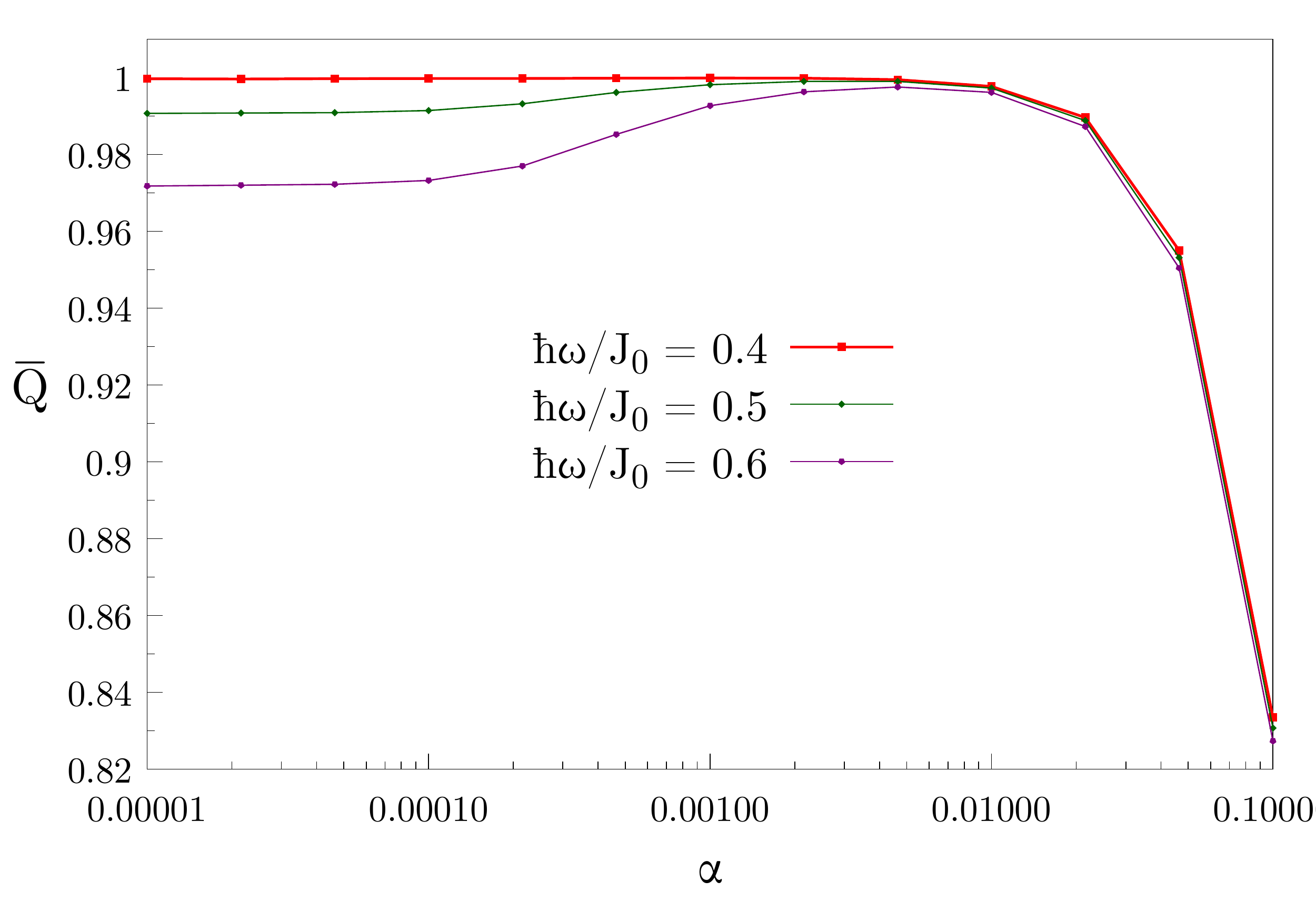}\\
        \end{tabular}
	\caption{Average pumped charge $\overline{Q}$ versus the coupling with the environment $\alpha$ for three different values of $\omega$ and $\tmpr=0$.}
	\label{avea:fig}
\end{figure}

We focus here on how the stationary pumped charge $\Qinfty$ changes as the interaction $\alpha$ is changed over different orders of magnitude. We have used the QME to study this behaviour and we show in Fig.~\ref{avea:fig} our results for $T=0$ and for the three frequencies $\hbar\omega/J_0 = 0.4, 0.5, 0.6$. 
Notice that we show results up to $\alpha = 0.1$ for completeness, but we consider our QME reliable up to $\alpha \simeq 10^{-2}$, see Ref.~\cite{Arceci_PRB17}. 
Data for $\hbar\omega \leq 0.4J_0$ are almost constant in the range $\alpha \in [10^{-5}, 10^{-2}]$ (we just show $\hbar\omega = 0.4J_0$). 
However, at larger frequencies things get more interesting: The curves are {\it non monotonic} and there is a maximum of $\overline{Q}$ at around $\alpha\simeq 6\times 10^{-3}$. 
There is therefore an interval of $\alpha$ where a stronger dissipation stabilizes the pumping against the driving. 

Focusing on the data in Fig.~\ref{avea:fig} for $\alpha = 0.001$,  
we can furthermore notice that the trend $\Qinfty = 1$ observed previously in Fig.~\ref{Figure1}(c) starts to deviate from the quantised value at frequencies $\hbar \omega \simeq 0.5 J_0$. 

%It is interesting to study the physical mechanism by which this phenomenon occurs. When $\alpha$ is increased we find that the pumped charge gets nearer to the quantised value because the dissipation pushes the system nearer to the %lowest-energy Floquet state discussed above. In order to show this fact, we consider the time-averaged fidelity of the asymptotic density matrix with the lowest-energy Floquet state
%
%\begin{equation}\label{fid:eqn}
% F_{\rm LE} = \frac{1}{\tau}\int_0^\tau\bra{\Psi_-(t)}\hat{\rho}_{\rm asy}(t)\ket{\Psi_-(t)}\,,
%^\end{equation}
%
%where $\ket{\Psi_-(t)}\prod_k\ket{\psi_+^k(t)}$ and $\hat{\rho}_{\rm asy}(t)$ is the asymptotical $\tau$-periodic density matrix. As we can see in the lower panel of Fig.~\ref{avea:fig}, $F_{\rm LE}$ has a maximum in $\alpha$ exactly where there is the maximum of $\Qinfty$. There, $F_{\rm LE}$ is nearest to 1, marking that $\hat{\rho}_{\rm asy}(t)$ is nearest to the lowest-energy Floquet state.
%------------------------------------------------
\section{Conclusions}	\label{sec:conclusions}
%------------------------------------------------
%
We analysed the role of dissipation from a somewhat idealised thermal environment --- coupling independent baths to each fermionic $k$-mode ---
on Thouless pumping in the Rice-Mele model. 
We found that a low temperature bath can assist against undesired (inevitable) non-adiabatic effects. 
Indeed, at fixed finite driving frequency, the pumped charge obtained from dissipative evolution can be closer to the quantised value with respect 
to the one obtained from purely coherent dynamics \cite{Privitera_PRL18}.  
Dissipation induces this improvement because it increases the population of the lowest-energy Floquet band. 
Indeed, the pumped charge would be essentially quantised --- up to exponentially small terms --- when this band is completely filled.
This is somewhat in line, in a non-topological context, with the finding of improved pumping in a three-site fermionic chain \cite{Pellegrini_PRL11} 
%\RED{and in a Cooper-pair sluice~\cite{Thingna_PRB14}, both }
subject to dissipation. 

Our findings are qualitatively independent of the system-bath coupling chosen as long as we stay in a weak coupling regime, as discussed in~\ref{App_couplSx}. 
They are furthermore validated by benchmarking against a numerically-exact method that does not rely upon any weak-coupling nor Born-Markov approximation. 
%
%\BLUE{I would remove: We also remark that the phenomena we see are qualitatively robust if we change the specific approximations behind the quantum master equation we use 
%(although the quantitative details are different), as we detail in~\ref{App_Eqs}.
%}

The fact that thermal effects can be beneficial is remarkable and interesting for future experimental realizations. 
We stress that the effect is not related to a {\em bath engineering}, exploited in the literature for other topological models\cite{Budich_PRA15}.
%However, this is not the first example of exploiting dissipation in order to achieve some specific purpose in a topologically non-trivial model. 
%In the setting of fermions hopping on a 2-dimensional lattice, Lindblad dissipators have been specifically engineered to prepare a Chern insulator in presence 
%of noise \cite{Budich_PRA15}. 
%However, we stress that the aim of the latter work is completely different from the present paper's one. 

A further step towards a deeper understanding would be to study more realistic couplings to the environment, \textit{e.g.} via operators 
acting on sites in real space, which break the entanglement in physical space. 
%In this way, the bath would no more affect each $k$-sector separately and 
% and this might lead to results different from the ones presented in this work. 
However, this analysis requires more sophisticated approaches \cite{Daley_AP14, Jaschke_QST18, Chin_JMP10}, and is left to future studies. 

Another interesting direction would be to study topological measures, such as the Uhlmann phase~\cite{Viyuela_PRL14} and the Ensemble Geometric Phase (EGP)~\cite{Bardyn_PRX18}. In particular, it would be interesting to inquire if the Uhlmann phase of the asymptotic time-periodic effective density matrix has a relation with the pumped charge, in analogy with the Berry or the Aharonov-Anandan geometric phase in the coherent cyclic case~\cite{Russomanno_PRB11,Thouless_PRB83}.

\ack
We acknowledge fruitful discussions with R. Fazio, S. Barbarino, L. Privitera and M. Wauters. 
Research was partly supported by EU FP7 under ERC-ULTRADISS, Grant Agreement No. 834402.
GES acknowledges that his research has been conducted within the framework of the Trieste Institute for Theoretical Quantum Technologies (TQT).

\vspace{3mm}

%%%%%%%%%%%%%%%%%%%
\appendix
%%%%%%%%%%%%%%%%%%%
\section{Dependence on the cutoff}	\label{AppRM_alpha}
%
%\begin{figure}
%\centering
%\includegraphics[width=0.7\columnwidth]{AlphaScaling_Tper20T1.pdf}
%\caption{Pumped charge after an infinite number of cycles $\Qinfty$ vs the bath interaction strength $\alpha$, for fixed $\period = 20 \hbar/J_0$ and $\tmpr = J_0$. 
%The arrow points to the value $\alpha = 0.001$, employed for all the results in this work.}
%\label{AppRM_AlphaScaling}
%\end{figure}
%
\begin{figure}
\centering
\includegraphics[width=0.7\columnwidth]{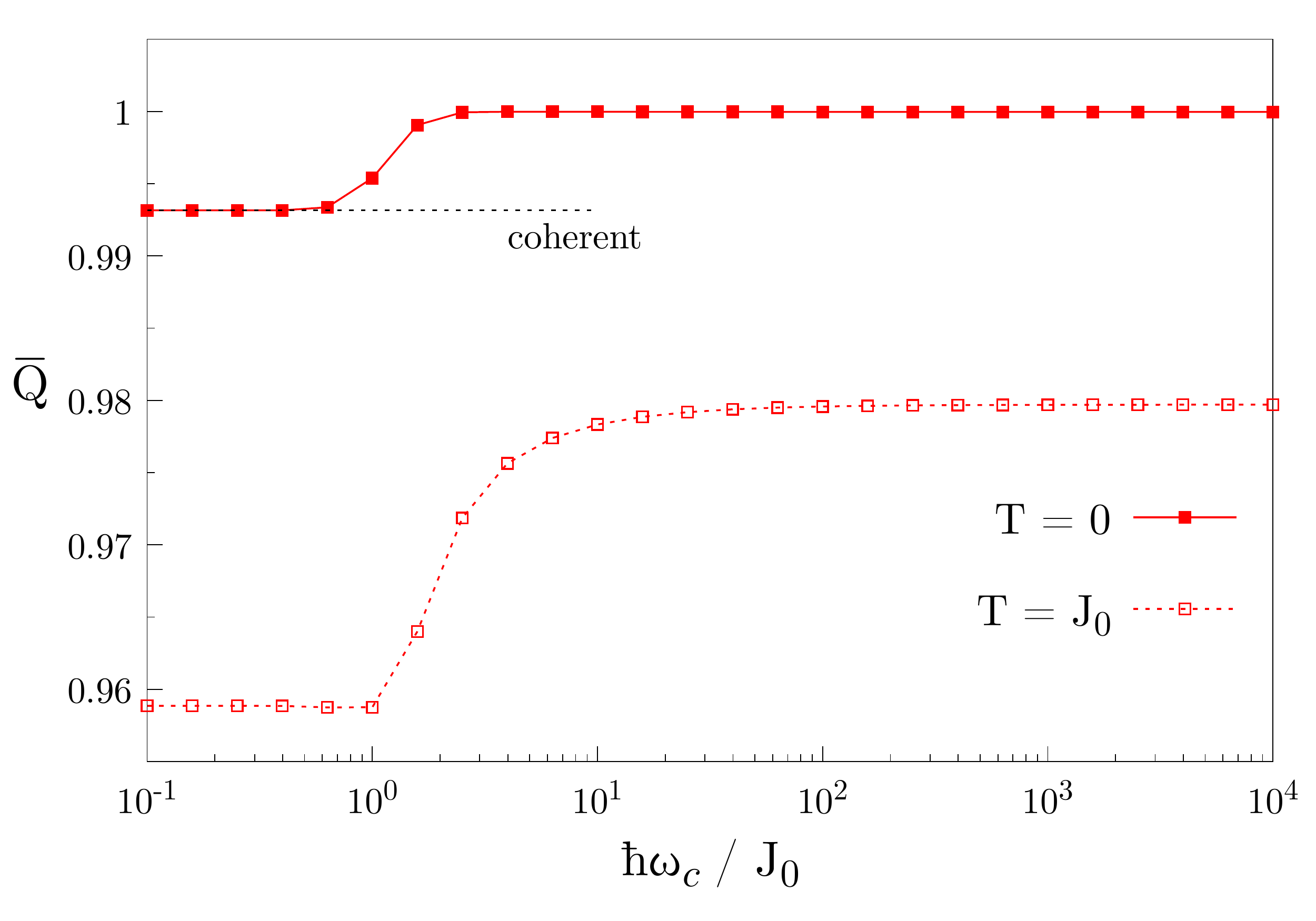}
\caption{$\Qinfty$ vs the cutoff frequency $\omega_c$ in the bath spectral function, at $\period = 20\;\hbar/J_0$ and at bath temperatures $\tmpr = 0, J_0$. 
The horizontal dashed line indicates the value of $\Qdiag$, the pumped charge at stationarity for the coherent dynamics, at the same $\period$.}
\label{AppRM_CutoffScaling}
\end{figure}
Let us focus on the issue of choosing the cutoff $\omega_c$ in the bath spectral function. 
Generally $\omega_c$ is taken to be the largest energy scale of the system, so that the dynamics becomes insensitive to the detail of this parameter. 
In the present case, since the system energy gap is always of the order of $J_0$ and we consider temperatures $\tmpr \leq J_0$, 
we require $\hbar \omega_c \gg J_0$. 
The behavior of $\Qinfty$ vs the cutoff $\omega_c$, see Fig.~\ref{AppRM_CutoffScaling}, 
shows the range of cutoff frequencies for which we observe a convergence of $\Qinfty$. 
We therefore selected $\hbar \omega_c = 1000 J_0$. 
Fig.~\ref{AppRM_CutoffScaling} is also useful to illustrate the effect of some basic dissipation mechanisms. 
If $\omega_c$ is much smaller than the minimum system energy gap, the probability of having jumps between energy levels is negligible and the result tends 
to become again insensitive to the cutoff value. Then, the only relevant dissipation mechanism comes from pure dephasing, 
given by $\Gpuredeph$ in Eq.~\eqref{PureDephasingRate}. Notice however that $\Gpuredeph \sim \tmpr$ and hence it vanishes at $\tmpr = 0$. 
This is consistent with what we observe in Fig.~\ref{AppRM_CutoffScaling}: for $\hbar \omega_c \ll J_0$, $\Qinfty$ is insensitive to the cutoff; 
moreover, for $\tmpr = 0$, we recover precisely the coherent result, pinpointed by the horizontal dashed line. 
%...............................................................................................................................................................

%++++++
\section{Different approximations in the Bloch-Redfield QME} \label{App_Eqs}
%++++++
%
In this appendix, starting again from the Bloch-Redfield QME, we derive and employ two sets of equations alternative to Eq.~\eqref{EqsToSolve} 
for the study of the steady state pumped charge. 
\begin{figure}
\centering
\includegraphics[width=0.7\columnwidth]{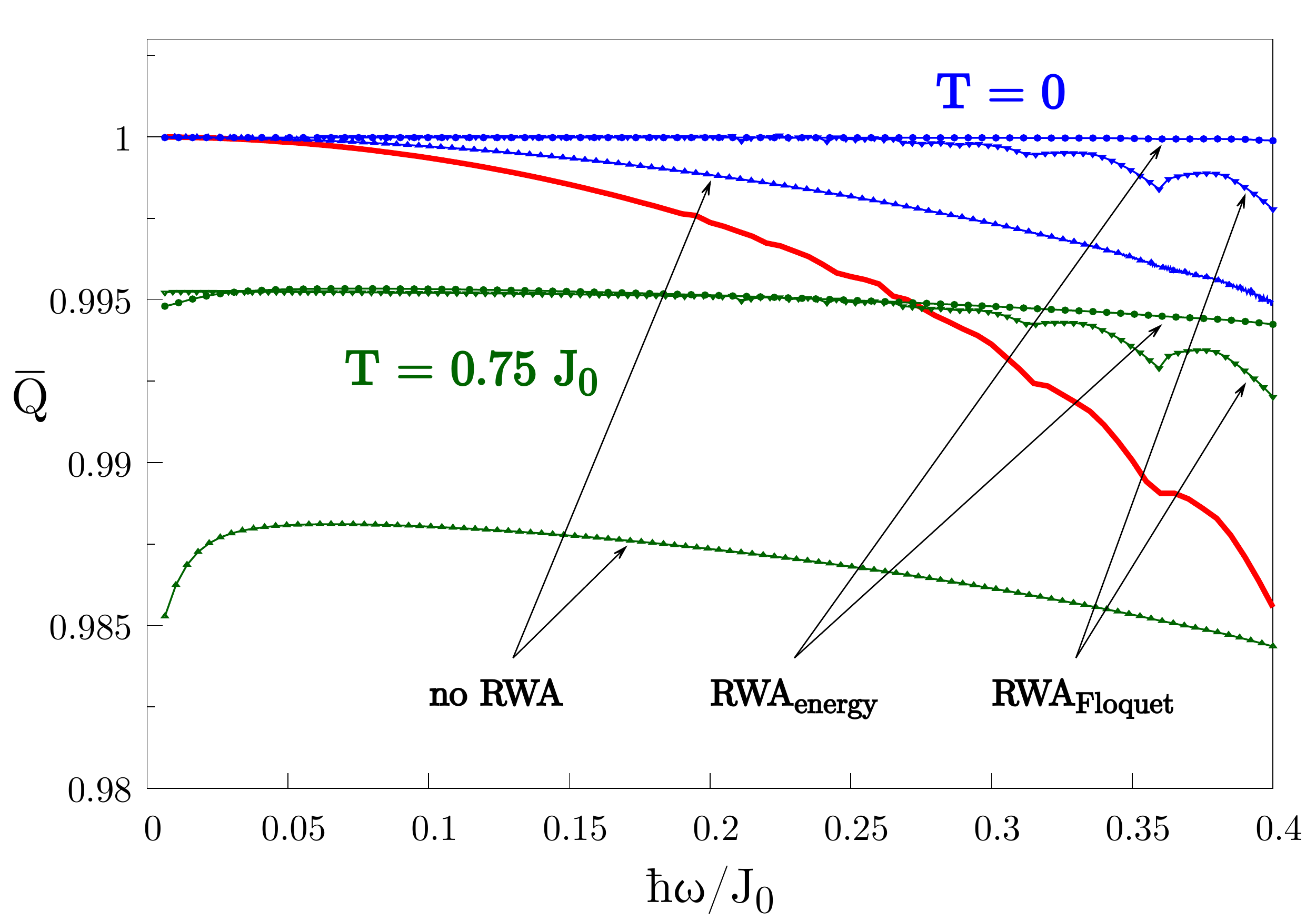}
\caption{$\Qinfty$ vs driving frequency $\omega$ for the dynamics induced by the Bloch-Redfield equation under different approximations. 
For all the cases, we still find a remarkable improvement over the coherent results if $\tmpr$ is low enough. 
Notice, in particular, the good agreement between the two different RWA schemes.}
\label{AppRM_approaches}
\end{figure}

The first one simply consists in the Bloch-Redfield QME \textit{without} any additional RWA. 
The equations have again the general form written in Eq.~\eqref{EqsToSolve}, with the coherent part unchanged, while the incoherent part, 
for $\n=(001)$ system-bath coupling, reads 
\begin{equation}	\label{Eqs_noRWA}
\tilde{\mathA}_{\mathrm{diss}} = \left( 
	\begin{array}{lll}
		\GdephNORWA & 0 & \gammaNORWA_{xz} \\
		0 & \GdephNORWA & \gammaNORWA_{yz} \\
		0 & 0 & 0
	\end{array} \right) \;,
\end{equation}
where 
\begin{subequations}
\begin{align}
	\GdephNORWA &= \Grelax + \Gpuredeph \\
	\gammaNORWA_{ij} &= \frac{R_i R_j}{\hbar^2 E^2} \big( S_X(2E/\hbar) - S_X(0) \big)
\end{align}
\end{subequations}
and the vector $\bfb$ is 
\begin{equation}	\label{Eqs_noRWA_const}
	\bfb = \frac{S_X(2E/\hbar)}{E \hbar^2} \tanh(\beta E) 
	\left( R_x, R_y, 0 \right)
\end{equation}

The second approach makes use of the Bloch-Redfield QME expanded in the coherent system Floquet basis, 
$\{ \ket{\Fstate_{n}^k(t)}\}_{{n} = \pm}$, defined by Eq.~\eqref{Floquet}. 
We will consider here a single dissipative two-level system at fixed momentum and we will omit the $k$ label in all the operators for clarity.
Following Refs.~\cite{Kohler_PRE97, Hausinger_PRA10, Russomanno_PRB11, Russomanno_2017}, it is possible to perform 
a RWA according to the quasi-energies (analogous to the standard one done on the system's energies). 
Eventually, one would see that the equations for the coherences decouple from the ones for the populations. 
It can be shown that the coherences go to zero after a finite time, so that the steady state is diagonal in the Floquet basis. 
The populations $\rho_{{n}{n}}(t) = \bra{\Fstate_{{n}}(t)} \rhosys(t) \ket{\Fstate_{{n}}(t)}$ 
can be determined by the rate equation~\cite{Kohler_PRE97, Hausinger_PRA10, Russomanno_PRB11, Russomanno_2017} 
\begin{equation}	\label{AppRM_RateEquation}
	\dot{\rho}_{--}(t) = 
	W_{+ \rightarrow -} \; \rho_{++}(t) - W_{- \rightarrow +} \; \rho_{--}(t) \; ,
\end{equation}
where $\rho_{++} = 1 - \rho_{--}$ and the rates are given by 
\begin{equation}
	W_{{n'} \rightarrow {n}} = \frac{1}{\hbar^2} \sum_l |A_{{n'} {n}, l}|^2 \gamma(\Delta_{{n'} {n}, l}) \; ,
\end{equation}
where we defined $A_{{n'} {n}, l}$ as the $l^{th}$ Fourier coefficient of the $\period$-periodic function 
$\bra{\Fmode_{n'}(t)} (\n \cdot \bsigma) \ket{\Fmode_{n}(t)}$, while 
$\gamma(\omega)$ is the Fourier transform of the bath correlation function and 
$\Delta_{{n'} {n}, l} = (\Fqe_{n'} - \Fqe_{n})/\hbar - 2\pi l / \period$.
The steady state is then very easily determined by setting Eq.~\eqref{AppRM_RateEquation} to zero, which leads to 
\begin{equation}
	\rho_{--} = \frac{W_{+ \rightarrow -}}{W_{+ \rightarrow -} + W_{- \rightarrow +}} \; .
\end{equation}

We employed both approaches to compute the pumped charge at stationarity $\Qinfty$ vs the driving frequency. 
These results are compared to the ones shown in the main text in Fig.~\ref{Figure1}(c) for the dynamics induced by Eq.~\eqref{EqsToSolve}. 
In Fig.~\ref{AppRM_approaches}, we provide the outcomes from the three approaches: 
In the plot, ``$\textrm{RWA}_{\textrm{energy}}$'' refers to the RWA according to the instantaneous system's energies, 
``$\textrm{no RWA}$'' points to the data obtained from the Bloch-Redfield QME without RWA, while 
``$\textrm{RWA}_{\textrm{Floquet}}$'' refers to the RWA performed on the QME written in the system's Floquet basis. 
We observe that the improvement over the coherent curve is obtained in all the three cases, giving further support to our claims. 
Furthermore, the results obtained from the two versions of the RWA seem to match quite well, especially at smaller frequencies. 
Nevertheless, the various approached give results which are quantitatively different. %especially the ones obtained without RWA. 
For example, at $\tmpr = 0.75 J_0$ and in the frequency range studied, one might get or not an improvement over the coherent case depending on the approach used. 
It is hard to say which version of the QME is best approximating the true dissipative time-evolution, until a careful benchmarking against the numerically exact MPS
approach is performed for an extensive set of parameters. This is left to a future study.  
%The benchmark in Sec.~\ref{sec:compare_QMEMPS} might seem to point towards the QME with RWA according to the instantaneous system's energies, but this might not be true. 
%To answer this question one would need a more detailed comparison among all the approaches at different regimes, which is beyond the scope of this work. 
%...............................................................................................................................................................

\section{Dependence on the coupling operator}
\label{App_couplSx}
\begin{figure}
\centering
\includegraphics[width=0.7\columnwidth]{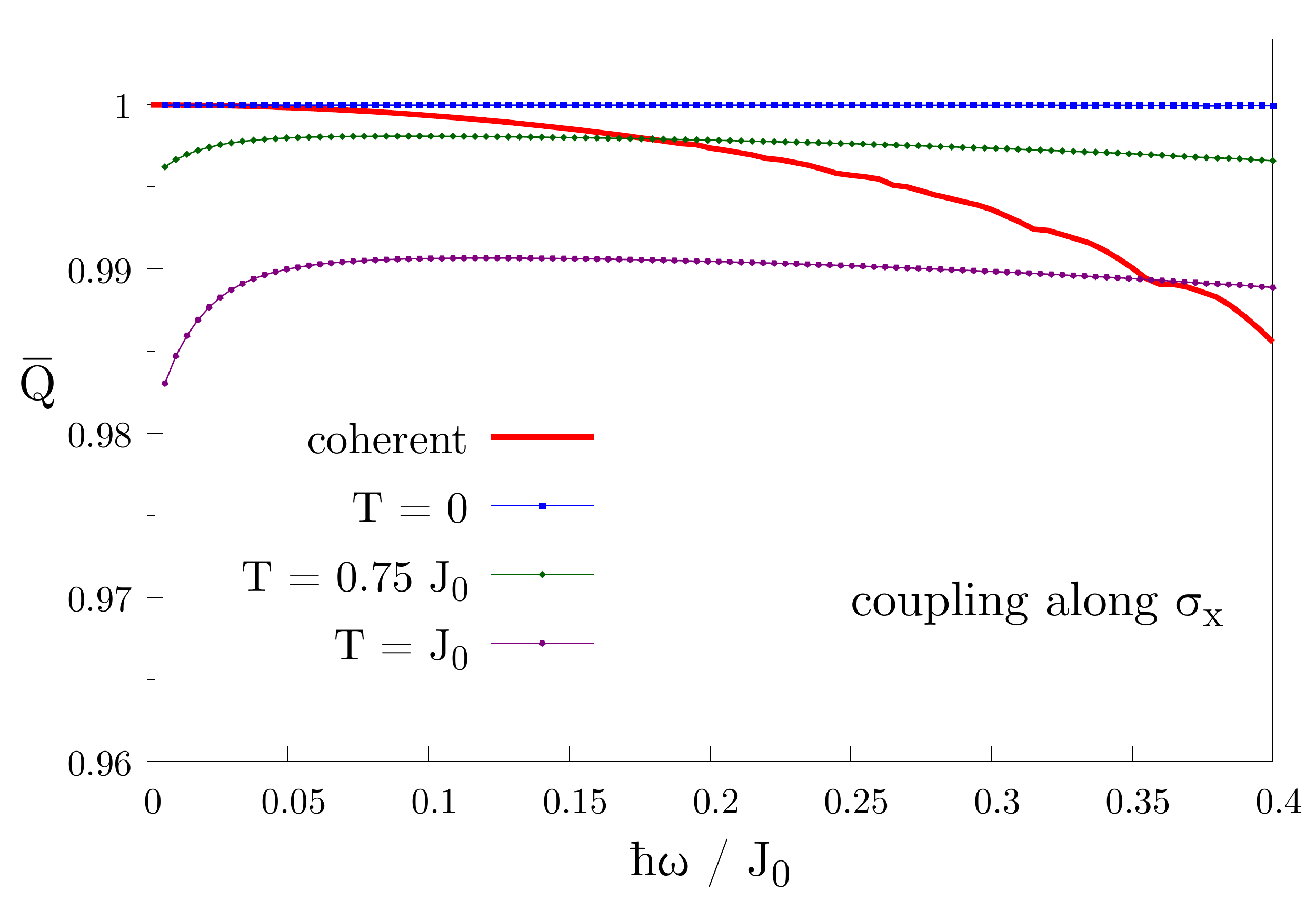}
\caption{$\Qinfty$ vs driving frequency $\omega$ for a system coupled to the bath via the $\pauli_x$ operator ($\n = (100)$). 
We observe a qualitative agreement with the result in Fig.~\ref{Figure1}(c) corresponding to a coupling along $\pauli_z$ ($\n = (001)$). }
\label{AppRM_CouplSigmaX}
\end{figure}
To test the generality of our findings, we study also system-bath coupling operators different from $\pauli_z$. 
We focus here on the case in which each two-level system is coupled to the reservoir via the $\pauli_x$ operator, which would correspond to choosing 
$\n = (100)$ in Eqs.~\eqref{PureDephasingRate} and \eqref{RelaxationRate}. 
Fig.~\ref{AppRM_CouplSigmaX} shows the result for $\Qinfty$ vs the driving frequency.
We see that there is no qualitative difference with the result shown in Fig.~\ref{Figure1}(c), corresponding to a coupling via $\pauli_z$. 
We tried also other coupling operators (for generic $\n$) and we obtained qualitatively similar results. 
\vspace{5mm}

\newpage
%\bibliography{../../Resubmission/BiblioRM,../../Resubmission/BiblioQIsingAngelo,../../Resubmission/BiblioMPS_OQS}

\begin{thebibliography}{10}
\bibitem{note_RWA}
The error is of the same order of the one coming from the discretization of $k$
  in the numerical evaluation of Eq.~\eqref{RM_Qdiag_diss}. If we use the RWA
  with respect to the quasienergies, as clarified in Sec.~\ref{App_Eqs},
  $\rhoFlqt{\alpha\alpha}(t)$ is time-independent and
  Eq.~\eqref{RM_Qdiag_diss_floquet} is exact in the framework of this RWA.
\end{thebibliography}

\end{document}